\documentclass[a12,preprint,onecolumn]{aastex}

\usepackage{graphicx}
\usepackage{amsmath}
\usepackage{multirow}
\usepackage{hyperref}
\usepackage{xspace}
\usepackage{url}

\renewcommand\arraystretch{1.2}

    \renewcommand{\arcsec}{\hbox{$^{\prime\prime}$}\xspace}
    \renewcommand{\degr}{\arcdeg\xspace}
    \newcommand{\kms}{\hbox{km s$^{-1}$}\xspace}
    \newcommand{\fluxcgs}{\hbox{erg s$^{-1}$ cm$^{-2}$}\xspace}
    \newcommand{\lumcgs}{\hbox{erg s$^{-1}$}\xspace}
    \newcommand{\xicgs}{\hbox{erg cm s$^{-1}$}\xspace}
    
    \newcommand{\ang}{\AA\xspace}
    \newcommand{\fek}{Fe\,K\xspace}
    \newcommand{\feka}{Fe\,K$\alpha$\xspace}
    \newcommand{\fekb}{Fe\,K$\beta$\xspace}
    \newcommand{\ka}{K$\alpha$\xspace}
    \newcommand{\kb}{K$\beta$\xspace}
    
    \newcommand{\hb}{H$\beta$\xspace}
    \newcommand{\fei}{Fe\,\textsc{i}\xspace}
    \newcommand{\fexxv}{Fe\,\textsc{xxv}\xspace}
    \newcommand{\fexxvi}{Fe\,\textsc{xxvi}\xspace}
    \newcommand{\oviii}{O\,\textsc{viii}\xspace}

    \newcommand{\dchi}{\Delta\chi^{2}}  
    \newcommand{\dchidnu}{\Delta\chi^{2}/\Delta\nu}
    \newcommand{\chidof}{\chi^{2}/\nu}

    \newcommand{\logxi}{\log\,(\xi/\rm{erg\,cm\,s}^{-1})}

    \newcommand{\lya}{Ly$\alpha$\xspace}

    \newcommand{\rg}{r_{\rm g}}
    \newcommand{\rgdef}{GM/c^2}
    \newcommand{\spin}{cJ/GM^2}
    
    \newcommand{\rmn}[1]{{\mathrm{#1}}}

    \newcommand{\suzaku}{\emph{Suzaku}\xspace}
    \newcommand{\xmm}{\emph{XMM--Newton}\xspace}

    \newcommand{\chandra}{\emph{Chandra}\xspace}
    \newcommand{\nustar}{\emph{NuSTAR}\xspace}
    \newcommand{\athena}{\emph{Athena}\xspace}
    \newcommand{\laor}{\texttt{laor}\xspace}
    \newcommand{\relline}{\texttt{relline}\xspace}
    \newcommand{\relxill}{\texttt{relxill}\xspace}

    \newcommand{\optxagn}{\texttt{optxagn}\xspace}

    \newcommand{\xspec}{\textsc{xspec}\xspace}
    \newcommand{\ciao}{\textsc{ciao}\xspace}

    \newcommand{\ark}{Ark\,120\xspace}
    
    \newcommand{\ngc}{NGC\,1365\xspace}
\shorttitle{\fek transients in \ark}
\shortauthors{E. Nardini et al.}

\begin{document}

\title{A deep X-ray view of the \textit{bare} AGN \ark. 
II. Evidence for \fek emission transients}

\author{E.~Nardini\altaffilmark{1}, D.~Porquet\altaffilmark{2}, 
J.~N.~Reeves\altaffilmark{1,3}, V.~Braito\altaffilmark{3,4}, 
A.~Lobban\altaffilmark{5}, and G.~Matt\altaffilmark{6}}

\altaffiltext{1}{Astrophysics Group, School of Physical and Geographical Sciences, 
Keele University, Keele, Staffordshire ST5 5BG, UK}
\altaffiltext{2}{Observatoire Astronomique de Strasbourg, CNRS, UMR 7550, 
11 rue de l'Universit\'e, 67000 Strasbourg, France}
\altaffiltext{3}{Department of Physics, University of Maryland Baltimore County, 
1000 Hilltop Circle, Baltimore, MD 21250, USA}
\altaffiltext{4}{INAF -- Osservatorio Astronomico di Brera, via E. Bianchi 46, 
I-23807 Merate, Italy}
\altaffiltext{5}{Department of Physics and Astronomy, University of Leicester, 
Leicester LE1 7RH, UK}
\altaffiltext{6}{Dipartimento di Matematica e Fisica, Universit\`a degli Studi 
Roma Tre, Via della Vasca Navale 84, I-00146 Roma, Italy}

\email{e.nardini@keele.ac.uk}

%%%%%%%%%%%%%%%%%%%%%%%%%%%%%%%%%%%%%%%%%%%%%%%%%%%%%%%%%%%%%%%%%

\begin{abstract}
We report on the results from a large observational campaign on the bare Seyfert galaxy \ark, 
jointly carried out in 2014 with \xmm, \chandra, and \nustar. The fortunate line of sight to 
this source, devoid of any significant absorbing material, provides an incomparably clean view 
to the nuclear regions of an active galaxy. Here we focus on the analysis of the iron fluorescence 
features, which form a composite emission pattern in the 6--7 keV band. The prominent \ka line 
from neutral iron at 6.4 keV is resolved in the \chandra High-Energy Transmission Grating spectrum 
to a full-width at half maximum of 4700$^{+2700}_{-1500}$ \kms, consistent with an origin from the 
optical broad-line region. Excess components are detected on both sides of the narrow \ka line: 
the red one (6.0--6.3 keV) clearly varies in strength in about one year, and hints at the presence 
of a broad, mildly asymmetric line from the accretion disk; the blue one (6.5--7.0 keV), instead, 
is likely a blend of different contributions, and appears to be constant when integrated over long 
enough exposures. However, the \fek excess emission map computed over the 7.5 days of the \xmm 
monitoring shows that both the red and the blue features are actually highly variable on timescales 
of $\sim$\,10--15 hours, suggesting that they might arise from short-lived hotspots on the disk 
surface, located within a few tens of gravitational radii from the central supermassive black hole 
and possibly illuminated by magnetic reconnection events. Any alternative explanation would still 
require a highly dynamic, inhomogeneous disk/coronal system, involving clumpiness and/or instability. 
\end{abstract}


%%%%%%%%%%%%%%%%%%%%%%%%%%%%%%%%%%%%%%%%%%%%%%%%%%%%%%%%%%%%%%%%%
\section{Introduction}

About four decades have passed since a crude shape of the X-ray continuum from active 
galactic nuclei (AGNs) was first revealed (e.g., Ives et al. 1976; Mushotzky et al. 1980), 
yet the origin of the main spectral components that have been progressively brought out 
is still largely unclear. The primary X-ray emission, usually described by a power law 
with high-energy ($\ga$\,100 keV) cutoff, is thought to stem from inverse Compton scattering 
of ultraviolet (UV) disk photons in a coronal region of hot electrons (e.g., Haardt \& 
Maraschi 1993). The nature of this corona and of its coupling with the disk, however, 
remains unknown. Neither do widely accepted explanations exist for the soft (Done et 
al. 2012; Vasudevan et al. 2014) or hard X-ray excesses (Risaliti et al. 2013; Miller 
\& Turner 2013), or the broad iron fluorescence lines at $\sim$\,5--7 keV (Tanaka et 
al. 1995; Inoue \& Matsumoto 2003). In fact, intricate absorption by neutral and/or ionized 
gas along the line of sight can reproduce any sort of spectral curvature or deviation from 
the power-law continuum. The unique potential of X-ray observations as a means of probing 
the immediate surroundings of an accreting supermassive black hole (SMBH) is therefore best 
exploited in the so-called \textit{bare} active galaxies, where any obscuration effect is 
negligible.

The nearby ($z \simeq 0.0327$; Osterbrock \& Phillips 1977) Seyfert galaxy \ark is arguably 
the most remarkable object in the bare AGN subclass. It is the X-ray brightest source of 
this type ($f_\rmn{0.3-10\,keV} \sim f_\rmn{14-195\,keV} \sim 7 \times 10^{-11}$ \fluxcgs; 
this work; Baumgartner et al. 2013), and it offers the cleanest view to the central engine. 
Early high-resolution data taken with the Reflection Grating Spectrometer (RGS) onboard \xmm 
posed stringent constraints on the presence of any warm absorber, whose column density 
would be at least an order of magnitude lower than found in a typical Seyfert 1 (Vaughan 
et al. 2004). Moreover, no evidence for UV absorption emerges from the \textit{Hubble 
Space Telescope} spectrum (Crenshaw et al. 1999). Since it is not affected by any 
significant foreground screen, the observed X-ray emission of \ark can thus be regarded 
as representative of the intrinsic high-energy output of an AGN, and of the physical 
processes associated with the inner accretion flow, such as Comptonization and/or 
relativistic reflection. \ark displays all the X-ray spectral traits expected for an 
unobscured, radiatively efficient SMBH, namely a smooth soft excess below 2 keV, an iron 
K-shell line complex possibly including a broad and skewed component, and a Compton 
reflection hump peaking at about 30 keV (e.g., Nardini et al. 2011). The soft excess 
has always been noticeable in \ark since the first X-ray observations (Brandt et al. 1993), 
but its featureless appearance is compatible with several interpretations, the most relevant 
of which invoke either the Comptonization of the seed UV photons in the warm disk atmosphere, 
or the blurring of the soft X-ray reflection-line forest in the strong gravity regime. The 
broad iron line and the Compton hump are even more puzzling, since their relative intensity 
apparently decreased in a recent (2013), combined observation with \xmm and \nustar, when 
the source was found to be two times fainter than usual (Matt et al. 2014).

In order to address these open issues and shed new light on the intrinsic X-ray emission 
of AGNs, \ark was the target of an extensive campaign carried out in 2014 March, consisting 
of an \xmm long look ($\sim$\,650 ks, or 7.5 days) with joint high-resolution view of the 
iron-K band with the \chandra High-Energy Transmission Grating (PI: D. Porquet), plus a 
simultaneous \nustar observation. Here we focus on the properties of the iron K-shell 
emission features, which can be studied in great detail thanks to the unprecedented depth 
of the new data sets. The paper is organized as follows: in Section~2 we introduce the 
various observations and provide the basic information on the data reduction, while Section~3 
concerns the analysis of the time-averaged spectra. Section~4 deals with the subsequent 
discovery of the short-term variability of iron fluorescence. These results and their 
possible implications on the physics of the X-ray corona are further discussed in Section~5, 
and conclusions are drawn in Section~6. The consequences of the soft X-ray, grating spectra 
on the bare character of \ark are the subject of a companion paper (Reeves et al. 2016), 
while the broadband analysis of the \xmm and \nustar observations, and the modeling of the 
optical to hard X-ray spectral energy distribution will be presented in a forthcoming work 
(D. Porquet et al. 2016, in preparation). 

%%%%%%%%%%%%%%%%%%%%%%%%%%%%%%%%%%%%%%%%%%%%%%%%%%%%%%%%%%%%%%%%%
\section{Observations and data reduction}

\ark was observed by \xmm over four consecutive orbits between 2014 March 18--24 (Table~\ref{to}). 
The event files were reprocessed with the Science Analysis System (\textsc{sas}) v14.0, applying 
the latest calibrations available on 2015 February. Due to the source brightness, the EPIC 
instruments were operated in Small Window mode, although this course of action was not enough 
to prevent pile-up in the MOS cameras. Only the EPIC/pn (Str{\"u}der et al. 2001) data were 
therefore taken into account, selecting the event patterns 0--4 (single and double pixel). 
The source spectra and light curves were extracted from circular regions centered on the target, 
with radius of 30\arcsec to avoid the edge of the chip. The background was evaluated over areas 
of the field of view where the contamination from the bright source is minimal, and its count 
rate is non negligible only in the final parts of each orbit, which were discarded. After the 
correction for dead time and background flaring, the total net exposure was $\sim$\,330 ks. 
Redistribution matrices and ancillary response files for all the data sets were generated with 
the \textsc{sas} tasks \texttt{rmfgen} and \texttt{arfgen}. Since there are several tens of 
counts per native energy channel in the \fek band (equivalent to a $\gg$\,5$\sigma$ significance 
at the observed background level), the 3--10 keV spectra were grouped into 30-eV bins not to 
unduly oversample the actual EPIC/pn resolution. This delivers about 1100 and 300 counts per 
bin around 5 and 8 keV, respectively. Due to some inaccuracy in the calibration of the EPIC/pn 
energy scale, a gain correction was also applied (see Appendix~A). 

As part of the 2014 campaign, \ark was also observed with the High-Energy Transmission Grating 
(HETG; Canizares et al. 2005) Spectrometer onboard \chandra, on March 17--22. This is the only 
\chandra observation of \ark to date. Due to scheduling constraints, it was split into three 
sequences with total exposure of $\sim$\,120 ks, overlapping with the two central \xmm orbits. 
The data were reprocessed with the \ciao v4.6 software package. Only the first-order spectra were 
considered for both the Medium- (MEG) and High-Energy Grating (HEG), and the $\pm1$ dispersion 
arms were combined. Only a modest ($\sim$\,10\%) flux variation was observed between the 
individual sequences, whose spectra were therefore merged into a single one for each grating. 
This yielded a total of $10.2 \times 10^4$ (MEG; 0.7--5 keV) and $4.7 \times 10^4$ (HEG; 1--8 keV) 
counts. The background contribution to the count rate was negligible. The resulting spectra were 
initially binned to $\Delta \lambda = 20$ (MEG) and 10 m\ang (HEG), which roughly correspond to 
the full width at half-maximum (FWHM) nominal resolution. To preserve the uniform spacing in 
wavelength units, in the HETG fits we made use of the $C$-statistic (Cash 1979), although the 
number of counts in each bin exceeds 25 over most of the range covered by the present analysis. 

In this study we also considered the two previous \xmm observations of \ark, taken respectively 
on 2003 August 24--25 (Vaughan et al. 2004) and on 2013 February 18--19 (again in combination 
with \nustar; Matt et al. 2014), as well as the \suzaku one, performed on 2007 April 1--3. The 
2003 and 2013 \xmm data were reduced with the same criteria adopted above.  The \suzaku spectra 
were re-extracted following the steps described in Nardini et al. (2011). For simplicity, here 
we used the single, merged spectrum from the two operating front-illuminated detectors (i.e., 
XIS\,0 and XIS\,3; Koyama et al. 2007), rebinned by a factor of four to 512 energy channels and 
further grouped to a minimum of 50 counts per bin. The observation log for all the data sets 
employed in this work is provided in Table~\ref{to}. The spectral analysis was performed using 
the \textsc{xspec} v12.9 package. All the fit uncertainties correspond to the 90\% confidence 
level ($\dchi$ or $\Delta C = 2.71$) for the single parameter of interest. Unless otherwise 
stated, the energy of the emission lines always refers to the rest frame of \ark. A concordance 
cosmology with $H_0=70$ km s$^{-1}$ Mpc$^{-1}$, $\Omega_m=0.27$, and $\Omega_\Lambda=0.73$ 
(Hinshaw et al. 2013) was assumed throughout.

\begin{table}
\centering
\small
\caption{Observation log of the X-ray spectra analyzed in this work.}
\label{to}
\begin{tabular}{c@{\hspace{15pt}}c@{\hspace{15pt}}c@{\hspace{15pt}}c@{\hspace{15pt}}c@{\hspace{15pt}}c}
\hline \hline
Mission & Obs.\,ID & Obs.\,Start (UTC) & Obs.\,End (UTC) & Exp.$^a$ & $\mathcal{C}^b$ (s$^{-1}$) \\
\hline
\multicolumn{6}{c}{\textit{2014 March campaign}} \\
\xmm & 0721600201 & 2014 Mar 18 -- 08:52:49 & 2014 Mar 19 -- 21:26:23 & 81,613 & 27.14$\pm$0.02 \\
\xmm & 0721600301 & 2014 Mar 20 -- 08:58:47 & 2014 Mar 21 -- 21:17:21 & 83,875 & 22.65$\pm$0.02 \\
\xmm & 0721600401 & 2014 Mar 22 -- 08:25:17 & 2014 Mar 23 -- 20:54:18 & 82,388 & 25.23$\pm$0.02 \\
\xmm & 0721600501 & 2014 Mar 24 -- 08:17:19 & 2014 Mar 25 -- 21:00:54 & 81,902 & 22.78$\pm$0.02 \\
\chandra & 16539 & 2014 Mar 17 -- 07:49:04 & 2014 Mar 18 -- 02:13:52 & 63,016 & 0.402$\pm$0.003 \\
\chandra & 15636 & 2014 Mar 21 -- 04:39:50 & 2014 Mar 21 -- 07:54:29 & 10,239 & 0.342$\pm$0.010 \\
\chandra & 16540 & 2014 Mar 22 -- 13:56:19 & 2014 Mar 23 -- 04:01:17 & 47,257 & 0.376$\pm$0.004 \\
\hline
\multicolumn{6}{c}{\textit{Previous observations}} \\
\xmm & 0147190101 & 2003 Aug 24 -- 05:35:43 & 2003 Aug 25 -- 12:44:33 & 78,166 & 26.37$\pm$0.02 \\
\xmm & 0693781501 & 2013 Feb 18 -- 11:39:53 & 2013 Feb 19 -- 23:54:11 & 87,721 & 10.33$\pm$0.01 \\
\suzaku & 702014010 & 2007 Apr 01 -- 18:07:26 & 2007 Apr 03 -- 21:43:25 & 100,864 & 2.186$\pm$0.003 \\
\hline
\end{tabular}
\flushleft
\small{\textit{Notes.} $^a$Net exposure in seconds. $^b$Source count rate over the 0.3--10 keV 
(\xmm/pn), 1--8 keV (\chandra/HEG), and 0.6--10 keV (\suzaku/XIS\,03) bands.} 
\end{table}

%%%%%%%%%%%%%%%%%%%%%%%%%%%%%%%%%%%%%%%%%%%%%%%%%%%%%%%%%%%%%%%%%
\section{Spectral analysis}

%%%%%%%%%%%
\subsection{\chandra/HEG spectrum}
\begin{figure}
\centering
\includegraphics[width=14cm]{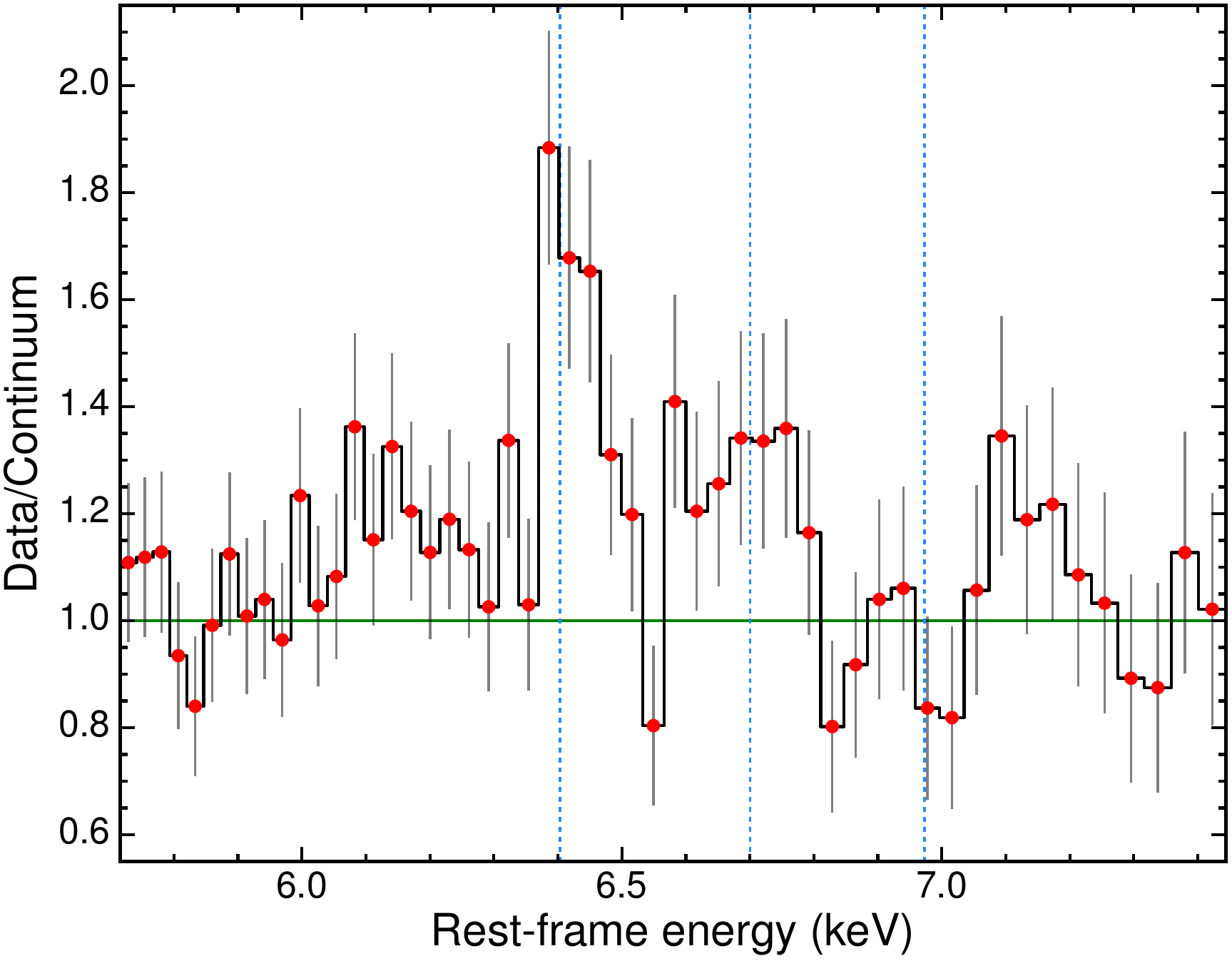}
\caption{Data to model ratio of the \chandra/HEG spectrum in the \fek band, obtained 
by fitting the continuum with the exclusion of the observed 5--7 keV range. A complex 
emission pattern clearly emerges, dominated by the narrow ($\sigma \simeq 40$ eV), 
resolved feature peaking around 6.4 keV. The vertical dashed lines correspond to the 
expected energies of the \fei, \fexxv, and \fexxvi \ka transitions. The data are 
binned to the FWHM resolution.}
\label{ch}
\end{figure}

We started our analysis from the \chandra observation, since the high-resolution data would 
allow us to determine the intrinsic width of any narrow \fek emission component. Here we 
concentrate on the 3--8 keV band of the HEG spectrum only, referring to Reeves et al. (2016) 
for a thorough discussion of the soft X-ray properties. We first fitted the continuum with a 
simple power law, after excluding the observed 5--7 keV energy range. A more complex model is 
not required, due to the absence of any spectral curvature above 3 keV (Appendix~B). This 
leaves clear residuals in emission at 6.0--7.2 keV in the rest frame, the most prominent of 
which is a narrow line at $\sim$\,6.4 keV (Figure~\ref{ch}). A pair of broader and shallower 
features are seen redwards and bluewards of this core, and a hint of a further line is present 
at $\sim$\,7.1 keV. Based on the visual inspection, we therefore included four Gaussian profiles, 
letting the centroid energy, width, and normalization free to vary for each of them. We also 
adopted the finer resolution of 5 m\ang (or 15 eV at 6 keV), which turns out to be more 
appropriate to fit such a rich and irregular \fek spectrum. This model gives a statistic of 
$C/\nu = 509/504$, and the best-fit line parameters are summarized in Table~\ref{tc}. For ease 
of illustration, at this stage we dub the three major lines as K$\alpha_\rmn{red}$, 
K$\alpha_\rmn{core}$, and K$\alpha_\rmn{blue}$, respectively; their actual nature will be 
discussed in more detail later on. The faintest line at $\sim$\,7.1 keV (marginally consistent 
with \fekb) is barely significant with its $\sim$\,25 counts (against $\sim$\,115 for 
K$\alpha_\rmn{core}$, $\sim$\,80 for K$\alpha_\rmn{red}$, and $\sim$\,65 for K$\alpha_\rmn{blue}$). 
Indeed, after its removal, the change in the fit statistic is completely negligible. As the 
HEG spectrum becomes rather noisy beyond 7 keV (with $<$\,10 counts per 5-m\ang bin), we do 
not consider this feature any further in this section. Of the three provisional \ka components, 
the red one is definitely the most puzzling, since it does not have the typical aspect of the 
gravitationally redshifted, extended wing of disk lines (e.g., Miller 2007). On the other hand, its 
energy of 6.13 keV and equivalent width (EW) of $\sim$\,55 eV also rule out the possibility that 
this is just the Compton shoulder (Yaqoob \& Murphy 2011, and references therein) of the \ka emission 
feature from neutral iron at 6.4 keV. Such an interpretation, in fact, is straightforward for the 
least ambiguous K$\alpha_\rmn{core}$ ($\rmn{EW} \sim 90$ eV), whose properties can thus be highly 
informative. Its width $\sigma_\rmn{core} = 43^{+22}_{-15}$ eV corresponds to a FWHM velocity 
broadening of 4700$^{+2700}_{-1500}$ \kms, in good agreement with the values obtained for \hb in 
the optical (5800--6100 \kms; Wandel et al. 1999; Marziani et al. 2003) and for the soft X-ray 
lines from He-like species (Reeves et al. 2016). The centroid of K$\alpha_\rmn{blue}$ at 6.68 keV, 
instead, would be compatible with a \fexxv \ka identification (but see below). Its slightly larger 
width compared to K$\alpha_\rmn{core}$ could be due to the emitting gas being closer to the 
illuminating source. However, a common velocity broadening of $\sim$\,5500 \kms, strikingly similar 
to the \hb one, returns an indistinguishable fit ($C/\nu = 510/505$). 

\begin{figure}
\centering
\includegraphics[width=14cm]{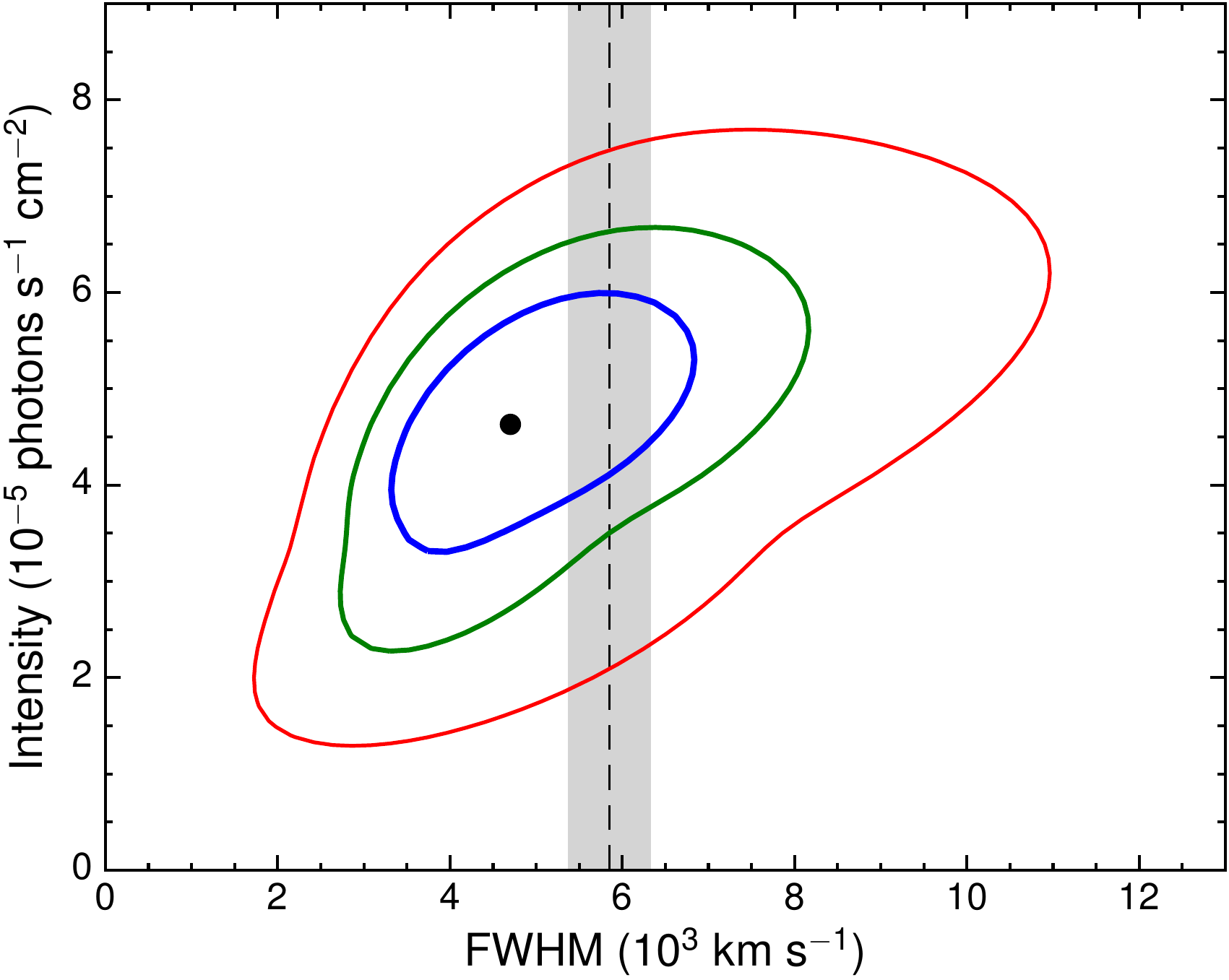}
\caption{Intensity versus velocity width contours at the nominal 68, 90, and 99\% 
confidence levels ($\Delta C = 2.3$, 4.61, and 9.21, respectively) for the \feka 
emission feature at $\sim$\,6.4 keV. Energy, width, and normalization of the red 
and blue components on each side of the core were allowed to vary. For comparison, 
the dashed line and the shaded area indicate the FWHM of the optical \hb line 
(5850$\pm$480 \kms; from Wandel et al. 1999).} 
\label{cc}
\end{figure}

We can safely exclude that the \ka emission core is unresolved; once we fix the width of 
K$\alpha_\rmn{core}$ to zero, the fit quality declines by $\Delta C/\Delta\nu \sim 11/1$. This 
alternative model is not satisfactory on qualitative grounds either, since both K$\alpha_\rmn{red}$ 
and K$\alpha_\rmn{blue}$ are shifted in energy and distorted into a blended, formless profile (compare 
with Figure~\ref{ch}). The confidence contours derived for K$\alpha_\rmn{core}$ are shown in 
Figure~\ref{cc}. In spite of some degeneracy with the adjacent features, the neutral \feka emission 
line is resolved at a $\sim$\,3$\sigma$ significance level. Its EW of 91$^{+28}_{-25}$ eV is broadly 
consistent with, yet somewhat larger than what expected in the context of the X-ray Baldwin effect 
(Iwasawa \& Taniguchi 1993). Considering the apparent anti-correlation of the \feka equivalent width 
with either the 2--10 keV X-ray luminosity ($\sim$\,10$^{44}$ \lumcgs in \ark) or the X-ray to 
Eddington luminosity ratio ($\sim$\,$5\times10^{-3}$), we should observe a value up to two times 
smaller (Bianchi et al. 2007; Shu et al. 2010). In principle, this leaves room for a hidden disk-line 
component, possibly asymmetric and contributing to some extent to both K$\alpha_\rmn{red}$ and 
K$\alpha_\rmn{blue}$. Indeed, it is worth noting that these structures can be excellently reproduced 
($C/\nu = 513/506$) as the red and blue horns of a disk line (e.g., Laor 1991) centered at 
6.41$\pm$0.05 keV, arising at a distance of $\sim$\,200 gravitational radii ($\rg = \rgdef$, where 
the black hole mass $M$ is $\sim$\,$1.5 \times 10^{8}\,M_{\sun}$ in \ark; Peterson et al. 2004; Ho 
\& Kim 2015) and seen at high inclination ($i \sim 80\degr$). An almost edge-on configuration would 
be hard to reconcile with the bare appearance of \ark. The geometry, however, is totally unconstrained 
($\Delta C = 1$ for $r_\rmn{in} = 6\,\rg$ and $i = 30\degr$). The narrow \ka core would still be 
resolved at $\sigma = 31^{+19}_{-12}$ eV, fully consistent with the value in Table~\ref{tc}. We then 
argue that a width of 40 eV is the most suitable for the subsequent analysis of the lower-resolution 
\xmm and \suzaku spectra. 

\begin{table}
\centering
\caption{\fek lines detected in the \chandra/HEG spectrum.}
\label{tc}
\begin{tabular}{l@{\hspace{20pt}}c@{\hspace{10pt}}c@{\hspace{10pt}}c@{\hspace{10pt}}c@{\hspace{10pt}}c}
\hline \hline
Line & $E$\,(keV) & $\sigma$\,(eV) & Norm. & EW\,(eV) & $\Delta C$ \\ 
\hline 
K$\alpha_\rmn{red}$ & 6.128$^{+0.121}_{-0.063}$ & 83$^{+317}_{-50}$ & 2.90$^{+4.97}_{-1.60}$ & 
53$^{+32}_{-31}$ & 14.0 \\ 
K$\alpha_\rmn{core}$ & 6.416$^{+0.016}_{-0.017}$ & 43$^{+22}_{-15}$ & 4.63$^{+1.56}_{-1.58}$ &
91$^{+28}_{-25}$ & 54.8 \\
K$\alpha_\rmn{blue}$ & 6.679$^{+0.044}_{-0.043}$ & 64$^{+47}_{-27}$ & 2.99$^{+1.63}_{-1.39}$ &
64$^{+32}_{-32}$ & 16.0 \\
K$\beta$ & 7.136$^{+0.077}_{-0.069}$ & $<$\,129 & 1.46$^{+1.48}_{-1.19}$ &
$<$\,72 & 4.3 \\
\hline
\end{tabular}
\flushleft
\small{\textit{Notes.} See the text for the adopted line nomenclature. Normalizations 
are in units of 10$^{-5}$ photons s$^{-1}$ cm$^{-2}$, and uncertainties are given at 
the nominal 90\% level ($\Delta C = 2.71$). The 3--8 keV continuum is modeled with 
a simple power law with $\Gamma = 1.91 \pm 0.07$. Since the lines are somewhat blended, 
EWs are computed with respect to the continuum only. The change in the statistic after 
the line removal ($\Delta C$, last column) is obtained without refitting to avoid 
meaningless solutions.}
\end{table}

%%%%%%%%%%%
\subsection{\xmm spectra (2014)}

\begin{figure}
\centering
\includegraphics[width=10cm]{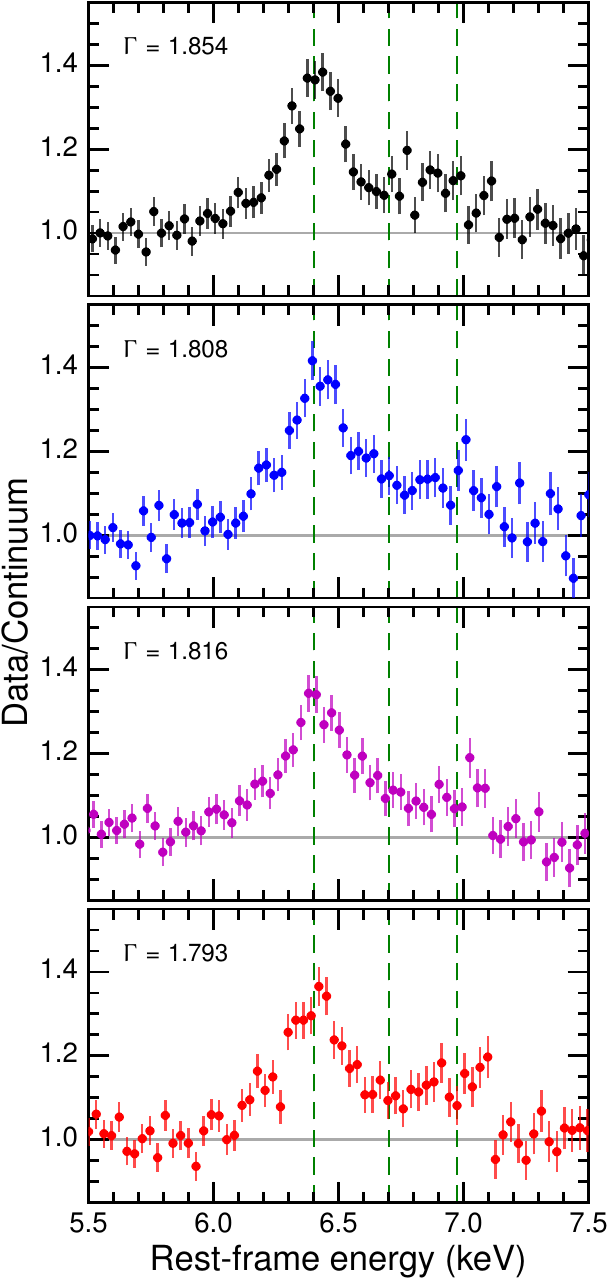}
\caption{\fek emission complex in the four consecutive \xmm observations of 2014 March, again 
obtained as the ratio between the data and the power-law continuum fitted at 3--5 plus 7--10 
keV (the best value of the photon index is reported in the left-hand top corner). Compared to 
the simultaneous, high-resolution \chandra spectrum, the 6.4-keV feature is of course much 
broader and less prominent, yet the red and blue excesses are still clearly present. Differently 
from the HEG case, the latter extends with no discontinuity up to at least 7 keV, not only due 
to the much better signal-to-noise of the \xmm spectra, but also to variability (see Section~4). 
The data are rebinned to a constant 30-eV resolution, and that the energies are gain-corrected.} 
\label{rx}
\end{figure}

\begin{figure}
\centering
\includegraphics[width=14cm]{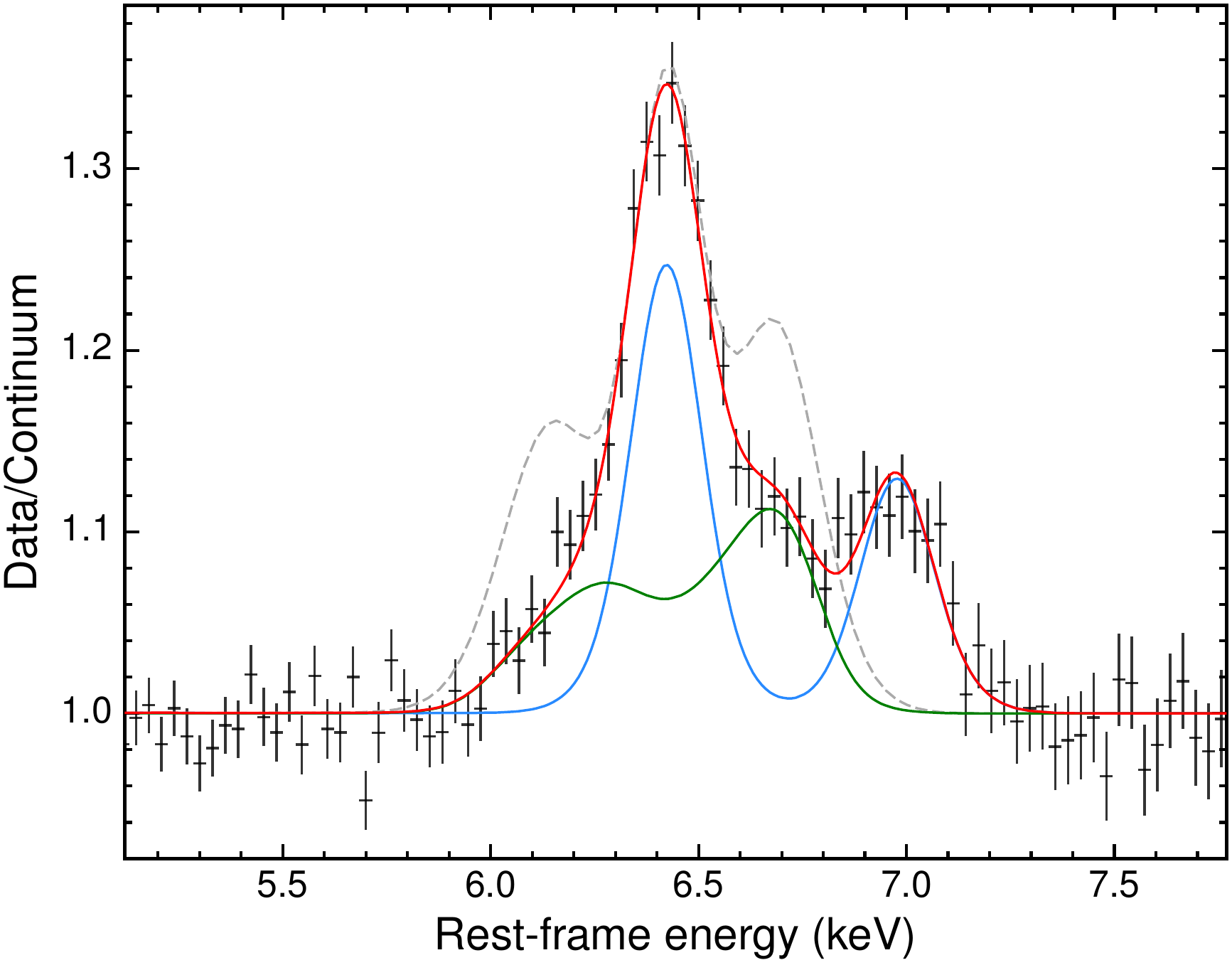}
\caption{Spectral decomposition of the \fek emission complex in the 2014$m$ \xmm 
spectrum (averaged over the entire campaign) when fitted with two narrow Gaussian 
lines at $E_1 \simeq 6.41$ keV and $E_2 \simeq 6.97$ keV (in turquoise) plus a 
relativistically distorted line from the mildly ionized accretion disk (in green). 
The latter feature apparently originates at several tens of gravitational radii 
from the SMBH, and cannot be statistically preferred to a very broad (FWHM $\sim$ 
30,000 \kms) Gaussian. The dashed line shows the overall profile of the three \ka 
components (red, core, and blue) in the \chandra/HEG best fit, which is a first 
hint of short-term variability.} 
\label{dl}
\end{figure}

As done with \chandra, we obtained an initial assessment of iron emission in the four 2014 \xmm 
spectra by ignoring the observed 5--7 keV band and fitting the remaining 3--10 keV continuum with a 
power law. The resulting data/model ratios are shown in Figure~\ref{rx}. Even if the overall \fek 
profile naturally looks smeared compared to the high-resolution HEG view, the presence of clear 
residuals on both sides of the main 6.4-keV line is fully confirmed. The most striking difference is 
that the blue excess seems to extend at least up to $\sim$\,7 keV, where no obvious feature is detected 
in the \chandra spectrum (Figure~\ref{ch}). This might be ascribed to the much larger EPIC/pn collecting 
area, which overrides the photon noise, but also short-term intensity fluctuations actually play an 
important role (see Section~4). Indeed, if we superimpose the HEG best fit on the \xmm data, only 
allowing for a cross normalization factor (of 0.7--0.8), the former K$\alpha_\rmn{red}$ and 
K$\alpha_\rmn{blue}$ are somewhat overestimated. Conversely, the strength of the core is perfectly 
matched, again suggesting that the shallower features are affected by noise and/or variability 
(Figure~\ref{dl}). We first ignored the findings from \chandra and tested a model simply consisting 
of a power law plus two broad Gaussian profiles (whose widths were tied as $\sigma_2 = 
E_2\,\sigma_1/E_1$). On statistical grounds, this fit would be already acceptable, returning a 
$\chi^2$ of 218, 264, 249, and 214 for 226 degrees of freedom (d.o.f.) in the four spectra (labeled 
as 2014a, -b, -c, -d hereafter). The implied FWHM is twice as large as the \hb one, and would call 
for an origin inside the optical broad-line region (BLR). However, this scenario is at odds with the 
high-resolution \chandra spectrum, whose availability then proves to be crucial for a sound \fek 
decomposition. 

We therefore resorted to the HEG-based information on the most likely intrinsic FWHM of the 6.4-keV 
\ka core, forcing the two lines to be substantially narrower ($\sigma_1 = 40$ eV). After re-fitting, 
the statistic in the single data sets deteriorates by $\dchi = 17$, 42, 32, and 25 for one extra 
degree of freedom (d.o.f.), respectively. We then included a third, broad Gaussian line, whose 
properties ($E_\rmn{G} \sim 6.45$ keV and $\sigma_\rmn{G} \sim 260$ eV on average; Table~\ref{tx}) 
do not vary significantly across the four spectra. Not only is this fit completely successful, but 
it is also marginally better than the initial, phenomenological one in each sequence, with $\chi^2$ 
now equal to 215, 258, 239, and 211 (for $\nu = 224$). As the energies and widths of all the lines 
are always consistent within the errors, we also analyzed the \textit{merged} spectrum (2014$m$, 
obtained by combining the four observations), in order to benefit from its higher quality. The 
statistical improvement yielded by the addition of a broad Gaussian over the initial, ill-informed 
model is definitely robust when referred to 2014$m$, with $\dchidnu = -21/-$2. Moreover, the 
position of the minor narrow peak moves from the unusual energy of 6.86--6.91 keV and is now 
virtually coincident with that of the \fexxvi \ka transition at 6.97 keV. The physical explanation 
for a line with a FWHM of $\sim$\,30,000 \kms is yet unclear. Applying the standard virial arguments, 
this emission feature should arise at a distance of $\sim$\,100\,$\rg$, which is the scale of the mid 
(outer) accretion disk, rather than that of the BLR.

Hence, it is worth considering an alternative model where the residual structures are described 
through a disk line, also given some previous claims based on the 2007 \suzaku observation 
(Nardini et al. 2011; Patrick et al. 2011; see also Vaughan et al. 2004 for earlier evidence). 
We tested several different relativistic line profiles, which all give the same results. Then we 
only focus on the latest and most sophisticated one, \relline (Dauser et al. 2010). With its EW 
$< 100$ eV, this feature is not as prominent as in some other, well-studied Seyfert galaxy. In 
\ngc, for instance, the broad \feka component might reach an EW of several hundreds eV, and also 
its overall flux is almost an order of magnitude greater than here (Risaliti et al. 2009). For this 
reason, not all of the line parameters in \ark can be well constrained simultaneously. We thus 
assumed no SMBH spin\footnote{The dimensionless spin parameter is defined as $a^* = \spin$, where 
$J$ and $M$ are the black hole's angular momentum and mass.}, a disc inclination of $i = 30\degr$, 
and an $\epsilon \propto r^{-q}$ emissivity law, with $q = 3$ (i.e., the Newtonian limit towards 
infinity). While the inclination conforms with the orientation of the host galaxy (26\degr; 
Nordgren et al. 1995), the other two conditions are physically consistent with the best-fit 
estimates for the disk inner radius, which is of the order of $r_\rmn{in} \sim 60\,\rg$ 
(Table~\ref{tx}). The rest-frame energy of the \relline component ($E_\rmn{R} \simeq 6.50$ keV) 
denotes a mildly ionized reflecting surface. Statistically, the `disk line' model is impossible 
to distinguish from the `broad Gaussian' one, with a difference of $\dchi = 1$ for the same 
number of d.o.f. (224) in 2014$m$. The spectral decomposition of the three lines is shown in 
Figure~\ref{dl}, where the \relline shape is only moderately skewed, since the effects of both 
gravitational redshift and relativistic beaming are rather weak at the large distances involved. 

\begin{table}
\centering
\caption{Fits to all the \xmm and \suzaku spectra of \ark in the 3--10 keV band.}
\label{tx}
\begin{tabular}{l@{\hspace{12pt}}c@{\hspace{8pt}}c@{\hspace{8pt}}c@{\hspace{8pt}}c@{\hspace{8pt}}c@{\hspace{8pt}}c@{\hspace{8pt}}c@{\hspace{8pt}}c}
\hline \hline
Obs. & 2014a & 2014b & 2014c & 2014d & 2014$m$ & 2003 & 2013 & 2007 \\ 
\hline 
\multicolumn{9}{c}{\texttt{powerlaw + 2 zgauss\,(narrow) + zgauss\,(broad)}} \\
$\Gamma$ & 1.84$^{+0.01}_{-0.01}$ & 1.80$^{+0.01}_{-0.02}$ & 1.80$^{+0.01}_{-0.01}$ &
1.78$^{+0.01}_{-0.01}$ & 1.79$^{+0.01}_{-0.01}$ & 1.83$^{+0.02}_{-0.01}$ & 
1.66$^{+0.02}_{-0.02}$ & 1.95$^{+0.02}_{-0.02}$ \\
$K$ & 1.13$^{+0.03}_{-0.02}$ & 0.96$^{+0.03}_{-0.02}$ & 1.10$^{+0.02}_{-0.03}$ & 
0.97$^{+0.03}_{-0.02}$ & 1.02$^{+0.01}_{-0.01}$ & 1.13$^{+0.03}_{-0.02}$ 
& 0.44$^{+0.02}_{-0.01}$ & 1.05$^{+0.04}_{-0.04}$ \\
$E_1$ & 6.39$^{+0.03}_{-0.01}$ & 6.40$^{+0.02}_{-0.02}$ & 6.40$^{+0.03}_{-0.03}$ & 
6.39$^{+0.02}_{-0.03}$ & 6.41$^{+0.01}_{-0.01}$ & 6.37$^{+0.03}_{-0.03}$ & 
6.40$^{+0.03}_{-0.03}$ & 6.40$^{+0.02}_{-0.01}$ \\
EW$_1$ & 65$^{+17}_{-17}$ & 56$^{+21}_{-20}$ & 39$^{+21}_{-18}$ & 52$^{+22}_{-22}$ & 
47$^{+10}_{-9}$ & 42$^{+18}_{-17}$ & 99$^{+57}_{-49}$ & 60$^{+16}_{-14}$ \\
$E_2$ & 6.90$^{+0.08}_{-0.08}$ & 6.98$^{+0.06}_{-0.05}$ & 7.00$^{+0.05}_{-0.06}$ & 
6.97$^{+0.04}_{-0.05}$ & 6.98$^{+0.03}_{-0.03}$ & 7.00$^{+0.04}_{-0.04}$ & 
7.00$^{+0.03}_{-0.04}$ & 6.96$^{+0.04}_{-0.03}$ \\
EW$_2$ & 20$^{+18}_{-16}$ & 25$^{+17}_{-16}$ & 20$^{+12}_{-12}$ & 30$^{+17}_{-14}$ & 
23$^{+7}_{-6}$ & 27$^{+11}_{-12}$ & 46$^{+12}_{-11}$ & 34$^{+16}_{-15}$ \\
$E_\rmn{G}$ & 6.46$^{+0.15}_{-0.16}$ & 6.48$^{+0.11}_{-0.05}$ & 6.42$^{+0.07}_{-0.06}$ & 
6.47$^{+0.13}_{-0.09}$ & 6.47$^{+0.05}_{-0.04}$ & 6.51$^{+0.09}_{-0.06}$ & 
6.61$^{+0.01}_{-0.18}$ & 6.36$^{+0.08}_{-0.09}$ \\
$\sigma_\rmn{G}$ & 290$^{+134}_{-157}$ & 263$^{+109}_{-86}$ & 251$^{+93}_{-70}$ & 
245$^{+105}_{-87}$ & 257$^{+50}_{-45}$ & 210$^{+73}_{-68}$ & $<$\,229 & 408$^{+103}_{-90}$ \\
EW$_\rmn{G}$ & 68$^{+28}_{-27}$ & 95$^{+24}_{-25}$ & 86$^{+22}_{-21}$ & 71$^{+24}_{-23}$ & 
86$^{+12}_{-11}$ & 83$^{+23}_{-22}$ & $<$\,88 & 164$^{+43}_{-40}$ \\
$\chidof$ & 215/224 & 258/224 & 239/224 & 211/224 & 260/224 & 251/224 & 271/224 & 208/219 \\
\hline
\multicolumn{9}{c}{\texttt{powerlaw + 2 zgauss\,(narrow) + relline}} \\
$\Gamma$ & 1.84$^{+0.01}_{-0.02}$ & 1.80$^{+0.01}_{-0.02}$ & 1.80$^{+0.01}_{-0.01}$ & 
1.78$^{+0.01}_{-0.01}$ & 1.79$^{+0.02}_{-0.01}$ & 1.83$^{+0.01}_{-0.01}$ & 
1.66$^{+0.02}_{-0.02}$ & 1.94$^{+0.03}_{-0.02}$ \\
$K$ & 1.13$^{+0.03}_{-0.02}$ & 0.96$^{+0.02}_{-0.02}$ & 1.10$^{+0.02}_{-0.03}$ & 
0.97$^{+0.02}_{-0.02}$ & 1.02$^{+0.01}_{-0.01}$ & 1.13$^{+0.03}_{-0.02}$ & 
0.44$^{+0.02}_{-0.01}$ & 1.04$^{+0.04}_{-0.03}$ \\
$E_1$ & 6.39$^{+0.02}_{-0.01}$ & 6.41$^{+0.01}_{-0.02}$ & 6.40$^{+0.01}_{-0.02}$ & 
6.39$^{+0.02}_{-0.01}$ & 6.41$^{+0.01}_{-0.01}$ & 6.40$^{+0.02}_{-0.02}$ & 
6.41$^{+0.03}_{-0.02}$ & 6.39$^{+0.02}_{-0.01}$ \\
EW$_1$ & 71$^{+11}_{-10}$ & 70$^{+13}_{-12}$ & 52$^{+13}_{-11}$ & 62$^{+12}_{-12}$ & 
58$^{+5}_{-5}$ & 48$^{+12}_{-12}$ & 79$^{+28}_{-23}$ & 70$^{+11}_{-11}$ \\
$E_2$ & 7.07$^{+0.13p}_{-0.37p}$ & 6.97$^{+0.05}_{-0.04}$ & 6.98$^{+0.05}_{-0.05}$ & 
6.97$^{+0.23p}_{-0.27p}$ & 6.97$^{+0.02}_{-0.03}$ & 6.99$^{+0.03}_{-0.03}$ & 
7.00$^{+0.03}_{-0.04}$ & 6.96$^{+0.03}_{-0.03}$ \\
EW$_2$ & $<$\,19 & 33$^{+9}_{-10}$ & 24$^{+8}_{-8}$ & 33$^{+11}_{-11}$ & 
29$^{+4}_{-4}$ & 33$^{+10}_{-10}$ & 48$^{+11}_{-11}$ & 50$^{+10}_{-10}$ \\
$E_\rmn{R}$ & 6.65$^{+0.09}_{-0.25}$ & 6.51$^{+0.07}_{-0.06}$ & 6.47$^{+0.05}_{-0.05}$ & 
6.54$^{+0.25}_{-0.10}$ & 6.50$^{+0.04}_{-0.03}$ & 6.52$^{+0.05}_{-0.03}$ & 
6.55$^{+0.07}_{-0.05}$ & 6.44$^{+0.05}_{-0.04}$ \\
$r_\rmn{in}$ & 37$^{+55}_{-13}$ & 50$^{+68}_{-19}$ & 58$^{+33}_{-21}$ & 60$^{+47}_{-33}$ & 
62$^{+16}_{-12}$ & 93$^{+60}_{-32}$ & 228$^{+72p}_{-130}$ & 24$^{+9}_{-13}$ \\
EW$_\rmn{R}$ & 83$^{+34}_{-33}$ & 78$^{+25}_{-20}$ & 71$^{+20}_{-17}$ & 61$^{+26}_{-19}$ &
69$^{+9}_{-9}$ & 72$^{+18}_{-15}$ & 63$^{+30}_{-23}$ & 142$^{+32}_{-32}$ \\
$\chidof$ & 213/224 & 259/224 & 238/224 & 209/224 & 261/224 & 249/224 & 270/224 & 206/219 \\
\hline
\end{tabular}
\flushleft
\small{\textit{Notes.} The power-law normalization ($K$) is given in units of 10$^{-2}$ 
photons keV$^{-1}$ s$^{-1}$ cm$^{-2}$ at 1 keV. Line energies are in keV, while widths 
($\sigma$ and EW) are in eV. The inner radius of the disk ($r_\rmn{in}$) is in gravitational 
radii. The `$p$' flag means that the uncertainty on a given parameter reaches the upper/lower 
limit allowed by the model. Black hole spin $a^* = 0$, disk inclination $i = 30$\degr, and 
emissivity index $q = 3$ are assumed for the \relline component.}
\vspace*{10pt}
\end{table} 

\subsection{Previous observations}

\begin{figure}
\centering
\includegraphics[width=16cm]{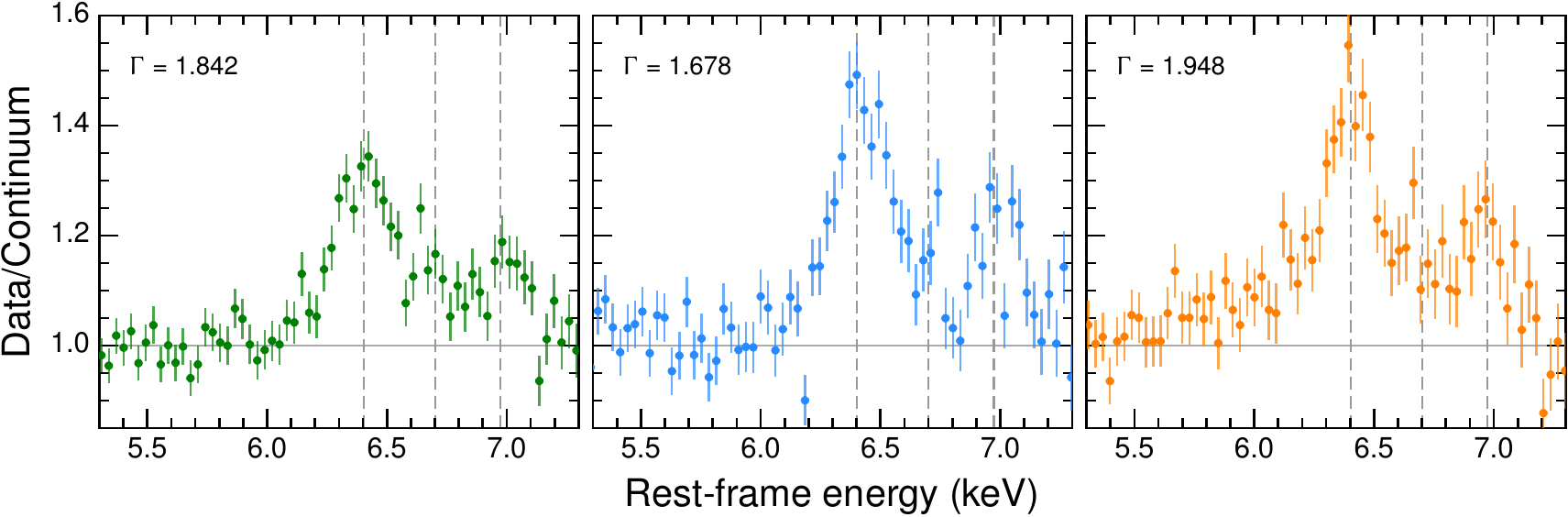}
\caption{Same as Figure~\ref{rx} for the 2003, 2013 \xmm, and 2007 \suzaku 
observations (from left to right). The red \fek wing is clearly variable in 
strength and extent on timescales of several years. Also the excess emission 
between the two narrow features is found at its lowest intensity alongside 
the wing, supporting their interpretation as signatures of disk reflection.} 
\label{ro}
\end{figure}

A broad \fek emission component is clearly involved in the 2014 \xmm spectra of \ark. In order 
to gain some more insights into its nature, we need to investigate its behavior over timescales 
of years. The analysis of the two previous \xmm observations of \ark is very useful in this 
sense. The spectrum of 2003 August, unfortunately, does not add much information. Indeed, its 
broadband shape and flux level are nearly identical to those observed in the first orbit of the 
2014 campaign down to $\sim$\,1 keV, where the soft excess exhibits a more pronounced roll-over 
compared to 2014a (Vaughan et al. 2004). Also the emission lines in the \fek band are in keeping 
with the best-fit parameters found about 11 years later in each model (see Table~\ref{tx} and 
Figure~\ref{ro}, left panel). The broad feature, in both the Gaussian and disk line cases, does 
not show any significant variation. Much more interesting is the 2013 observation (performed only 
13 months before the large campaign), as this is the only one so far that has caught \ark in a 
markedly different flux and spectral state (Matt et al. 2014). The 3--10 keV continuum is fainter 
by a factor of $\sim$\,1.7--2 with respect to all the other epochs, flatter ($\Gamma \sim 1.7$), 
and vaguely more \textit{convex}, with no obvious break energy between the soft excess and the 
hard power law. On the other hand, the relative strength of the former is very close to the 
average, once the different X-ray slope is taken into account. Following the continuum decrease, 
the EW of the two narrow lines grows by a similar factor (consistent within the uncertainties; 
Table~\ref{tx}), suggesting that there is no response on short ($<$\,1 day) timescales, comparable 
to the observation length. The broad/disk line, instead, clearly deviates from the above trend, 
since its EW is, if anything, smaller than in the 2003 and 2014 high-flux states. There is no 
compelling evidence for a red wing to the 6.4-keV feature, presumably because it is too faint, nor for 
any other kind of excess (Figure~\ref{ro}, middle panel). An additional Gaussian is therefore 
rather narrow ($\sigma \sim 80$ eV, although the upper limit is poorly constrained), and simply 
fits some minor residuals at 6.6--6.7 keV. Such a drastic width change from 2013 to 2014 is 
dubious, leaving once more the interpretation in terms of a (variable) fluorescence feature from 
the accretion disk as the most convincing one. 

The \relline component in the 2013 spectrum falls at a slightly higher energy ($E_\rmn{R} \simeq 
6.55$ keV), and its EW of $\sim$\,65 eV is comparable to the 2014$m$ one. This means that its 
intensity is roughly subject to the same attenuation experienced by the continuum. Other than 
this, the main discrepancy involves the disk inner radius, which moves much further out 
($r_\rmn{in} \ga 100\,\rg$). This is mostly due to the lack of a strong red wing, which removes 
the necessity of any asymmetry in the line profile, so tending to a Newtonian regime. Nevertheless, 
if both $E_\rmn{R}$ and $r_\rmn{in}$ are fixed to the values of 2014$m$, the fit statistic worsens 
by only $\dchidnu = 5/2$. While the ionization state of the gas, hence the line energy, can 
reasonably differ from 2013 to 2014, it is not clear which mechanism could be responsible for 
a recession of the disk inner boundary. In any case, the broad \feka component detected in \ark 
is apparently characterized by sizeable flux variability on timescales of just one year. 

The \suzaku observation performed in 2007 April conveniently covers the monitoring gap between 
the \xmm visits, which are separated by either one or ten years. Moreover, the recorded 3--10 
keV power-law flux of $\sim$\,$2.2 \times 10^{-11}$ \fluxcgs is almost exactly intermediate 
between the highest (2014c) and the lowest (2013) states, whereas the photon index is steeper 
than usual ($\Gamma = 1.95$)\footnote{Differently from the change in the photon index from 2013 
to 2014, this might not be entirely physical, given the cross-calibration systematics between 
the \xmm/EPIC and \suzaku/XIS instruments (e.g., Tsujimoto et al. 2011).}. We thus applied the 
same emission-line models employed in the \xmm analysis to the 3--10 keV \suzaku spectrum. The 
first outcome is that the red excess has a much more appreciable wing-like form, extending down 
to at least 5.7 keV (Figure~\ref{ro}, right panel). This, in turn, implies that to reproduce this 
structure with a Gaussian profile (centered at lower energy, $E_\rmn{G} = 6.36^{+0.08}_{-0.09}$ 
keV) a width of $\sigma_\rmn{G} \simeq 410$ eV (FWHM $\sim$ 45,000 \kms) is required, which is 
probably unphysical unless associated to a face-on disk. Despite the good fit quality ($\chidof 
= 208/219$), this is not a viable alternative to the `disk line' model. The resort to a \relline 
component ($a^* = 0$, $i = 30$\degr, $q = 3$) yet brings a further improvement ($\dchi = -2$ for 
the same d.o.f.). As expected from the noticeable footprint of gravitational redshift, the disk 
inner radius moves closer to the innermost stable circular orbit, being about a factor of two 
smaller than in the 2014 \xmm observation ($r_\rmn{in} \sim 25$ versus $60\,\rg$). Most notably, 
while the EW of two narrow lines at 6.39 and 6.96 keV modestly grows following the established 
anti-correlation with the underlying flux, the EW of the disk line remarkably increases to 
$\sim$140\,($\pm$30) eV. This is twice as large as seen in 2014, against a 20--25\% diminution 
of the continuum. All the lines' parameters are listed in the last column of Table~\ref{tx}, 
and are found to be in agreement with the results of Nardini et al. (2011), where a \laor kernel 
(Laor 1991) with $i = 40$\degr was used instead. The \suzaku spectrum then delivers a strong 
confirmation of a disk line, whose behavior is largely unpredictable when compared to the overall 
hard X-ray intensity. 

The main results of the analysis of the best-quality, time-averaged spectra of \ark can be 
therefore summarized as follows:

\begin{itemize}
\item The \fek emission-line complex in \ark displays a rather composite structure, dominated 
in prominence by the narrow \ka feature from neutral iron peaking at 6.4 keV, which is resolved 
in the \chandra/HEG spectrum. The width value of $\sigma = 43^{+22}_{-15}$ eV (FWHM $\sim$ 4700 
\kms) is commensurate with the velocity broadening of \hb in the optical. On both sides of the 
main core, residual emission is detected in every epoch. In particular, the red, wing-like 
excess is not compatible with a \ka Compton shoulder in terms of energy and equivalent width, 
thus calling for the presence of a broad and shallow independent feature. 
\item At the CCD-resolution, yet much higher signal-to-noise of the \xmm and \suzaku spectra, a 
second narrow line is regularly identified around 7 keV, which would be temptingly attributed to 
\fexxvi. Indeed, the equivalent width of both the 6.4- and 7-keV features fairly anti-correlates 
with the continuum flux over a decade (2003 to 2014), implying that their absolute strength is 
approximately constant, or subject to modest variations at most. Hence these lines appear to be 
sensitive to the average continuum over long (months to years) timescales only, in keeping with 
a possible broad-line region origin as suggested by their width.
\item The additional residuals, which include the red wing down to 5.7--6 keV as well as a sort 
of plateau between the two narrow lines, account for up to $\sim$\,50\% of the total \fek emission. 
These components can be jointly fitted through a broad profile, either symmetric ($\sigma \sim 
200$--300 eV) or slightly skewed, as for a disk line emanating from a mild gravitational regime. 
The latter description is preferred on physical grounds, given the evidence of significant 
changes in both flux and shape over a period of just one year, which place this feature within 
several tens to a few hundreds of gravitational radii from the central black hole. 
\end{itemize}

The disk line, even if inconclusive for a precise measurement of the SMBH spin, has a valuable  
diagnostic potential in the study of the central regions of \ark. In the next section we build on 
the results above and explore the source of the primary X-ray emission through the disk reflection 
signatures. 

\subsection{Reflection geometry and the X-ray source}

The broad \fek feature in \ark is apparently emitted from a region where relativistic effects 
are mild. A disk truncation would be anomalous in a source with such an X-ray brightness. Indeed, 
by taking a disk extended down to the innermost stable circular orbit of a SMBH with maximal spin 
($a^* = 0.998$, $r_\rmn{in} = 1.235\,\rg$), the fit of the 2014$m$ spectrum gets worse but not 
so dramatically, provided that the exponent of the emissivity law is allowed to adjust to $q = 
1.4^{+0.3}_{-0.4}$. A flatter emissivity would suggest that the X-ray source is either diffuse 
or, in the lamp-post approximation, located fairly high above the disk, so that the light bending 
is not so extreme as to focus most of the illuminating radiation towards the very inner regions 
(Miniutti \& Fabian 2004; Dauser et al. 2013). In general, reflection off the surface of the 
accretion disk can be a powerful clue to the structure and properties of the irradiating source. 
With \textit{reflection} are designated all the effects of the interaction between X-rays and 
\textit{cold}, optically-thick matter (e.g., George \& Fabian 1991; Ross \& Fabian 1993), of 
which fluorescence from iron and the most abundant lighter elements is a key ingredient. In 
particular, the combination of photoelectric absorption and Compton backscattering imparts 
considerable curvature to the incident continuum. To take this into account, we replaced the 
power law and the disk line in our previous models with a \relxill spectrum (v0.4a; Garc\'ia et 
al. 2014), which encompasses both the direct and the reflected components in a self-consistent 
way. We initially maintained the same set of assumptions adopted for \relline (i.e., $a^* = 0$, 
$i = 30$\degr, and $q = 3$), and restricted again the analysis to the 3--10 keV energy range. This 
allowed us to avail ourselves of a decent continuum window, and yet avoid any heavy contamination 
from the soft/hard excess. All the seven spectra (six from \xmm plus one from \suzaku) were fitted 
together tying the energies of the two narrow features, as this is fully justified by the data 
(whereby the most general case has a probability of chance improvement of 0.48 according to an 
$F$-test). The outcome is very good ($\chidof = 1632/1575$), and provides some interesting 
indications. The inner radius of the disk is exactly the same as before ($r_\rmn{in} \sim 25\,\rg$; 
Table~\ref{tr}) for the \suzaku observation, when the broad \fek line was stronger. In all the 
other cases it tends to be higher than the values of Table~\ref{tx} by a factor of $\sim$\,2, but 
still consistent due to the larger uncertainties. In absolute terms, then, $r_\rmn{in}$ is always 
at great distance from the innermost stable orbit, irrespective of the SMBH spin. 

The two other basic variables of the new model are the ionization parameter (defined as $\xi = 
L/nr^2$, where $L$ is the ionizing luminosity, $n$ the density of the gas, and $r$ its distance 
from the source), which mainly determines the position of the line, and the reflection fraction 
$\mathcal{R}$, measured as the ratio between the amount of radiation impinging on the disk and 
that directly escaping towards infinity (e.g., Dauser et al. 2016). For both quantities the 
confidence intervals largely overlap across the different epochs, with $\logxi$ ranging from 
2.30$^{+0.30}_{-0.24}$ (2007) to 2.72$^{+0.18}_{-0.42}$ (2014b), and $\mathcal{R}$ ranging from 
0.20$^{+0.12}_{-0.06}$ (2014d) to 0.33$^{+0.06}_{-0.09}$ (2014c), except for the \suzaku spectrum 
where $\mathcal{R} = 0.56^{+0.14}_{-0.16}$ (Table~\ref{tr}). No correlation is then found between 
the hard X-ray flux and the ionization state, for which the most common species of iron are 
Fe\,\textsc{xix}--\textsc{xxv} (Kallman et al. 2004). While an exhaustive discussion of the 
broadband X-ray emission of \ark, including the \nustar data, is deferred to a forthcoming 
companion paper (Porquet et al. 2016), we anticipate here that the estimated reflection fraction 
is far too low to reproduce the observed soft excess. Reflection could still heavily contribute 
towards the flux rise below 2 keV, but this would require distinct properties from those inferred 
from \fek fluorescence. Indeed, some caution should be exercised when fitting the whole X-ray 
spectrum with a reflection model (e.g., Nardini et al. 2012; Walton et al. 2013), since the 
statistical weight of the broad iron line becomes negligible and the relativistic blurring can 
be driven to an extreme degree by the smoothness of the continuum\footnote{This can also be 
deduced from the comparison between the \relxill- and the \relline-based (Section 4.1) models
at 5--7 keV, the former being worse by $\dchi = 20$ for the same degrees of freedom: $\chidof = 
435/427$ against 415/427 without refitting. Even the tiniest curvature of the continuum (which 
is indeed better fitted by \relxill at 3--10 keV; see also Appendix~B) is followed at the expense 
of the line profile.}. This could hinder our ability to distinguish between reflection components 
arising from different zones of the disk. 

Both the disk emissivity pattern and the reflection fraction depend on the geometry, size, and 
location of the X-ray corona (Wilkins \& Fabian 2012; Dauser et al. 2013). We have so far 
assumed an emissivity index $q = 3$, which is the canonical value at large distance from an 
isotropic, pointlike source. As the size of the X-ray emitting region in AGNs is apparently of 
a few gravitational radii (Mosquera et al. 2013, and references therein), and $r_\rmn{in} \gg \rg$ 
in our fits, there are no self-consistency issues. However, we can reverse the approach, choosing 
the source configuration with its incorporated $\epsilon(r)$ and $\mathcal{R}$ first. The simplest 
and most widely explored approximation is the lamp-post geometry (Matt et al. 1991), for which we 
used the most recent \texttt{relxilllp} calculations.\footnote{\url
{http://www.sternwarte.uni-erlangen.de/~dauser/research/relxill/index.html}} The height of the 
source ($h$) along the black hole's rotation axis is the new free parameter of the model, which 
controls the illumination of the accretion disk, hence its emissivity and reflection fraction 
(through the \texttt{fixReflFrac = 1} option). Also the lamp-post scenario turns out to be 
statistically successful, with $\chidof = 1638/1575$. While the best-fitting $\xi$ values change 
by $\sim$\,10--20\% at most, $h$ and $r_\rmn{in}$ are generally comparable and poorly constrained 
at several tens of $\rg$ (Table~\ref{tr}). Such a height above the disk, although not affected by 
the computational shortcomings of \texttt{relxilllp} recently noted by Nied{\'z}wiecki et al. 
(2016), is much more difficult to explain than the nominal inner radius. The latter, in fact, 
does not entail the truncation of the accretion flow well before the last stable orbit. If the 
corona is radially extended over the inner disk, the reflection component can be completely 
smeared through Comptonization even for ordinary temperatures and optical depths (Wilkins \& 
Gallo 2015). Only reflection from the outer regions would be therefore discernible from the 
primary continuum. Conversely, X-ray reverberation studies of AGNs suggest that the source lies 
at only a few $\rg$ from the disk (Emmanoulopoulos et al. 2014). Fixing $r_\rmn{in}$ to 6\,$\rg$ 
results in a $\dchidnu = 7/7$, but pushes the source further up (with a lower limit of 
$h > 125$--230\,$\rg$ in the \xmm spectra; Table~\ref{tr}), while fixing $h$ to 10\,$\rg$ has 
appreciable drawbacks ($\dchidnu = 59/7$; chance probability of $1.2 \times 10^{-9}$). Based on 
the observed disk line, a lamp-post, on-axis configuration for the X-ray corona then appears 
unlikely for the unusual height. A careful examination of the subtle \fek spectral variations 
over different timescales (from several years down to less than a day) is therefore needed to 
pinpoint the location of the reflecting gas and of its source of illumination, which is not 
necessarily one and the same with the source of the hard X-ray continuum.

\begin{table}
\centering
\caption{Key parameters of the disk reflection models.}
\label{tr}
\begin{tabular}{l@{\hspace{15pt}}c@{\hspace{10pt}}c@{\hspace{10pt}}c@{\hspace{10pt}}c@{\hspace{10pt}}c@{\hspace{10pt}}c@{\hspace{10pt}}c}
\hline \hline
Obs. & 2003 & 2007 & 2013 & 2014a & 2014b & 2014c & 2014d \\ 
\hline
\multicolumn{8}{c}{\texttt{relxill (general)}} \\
$\Gamma$ & 1.85$^{+0.02}_{-0.02}$ & 1.99$^{+0.04}_{-0.03}$ & 1.67$^{+0.02}_{-0.02}$ & 
1.85$^{+0.02}_{-0.02}$ & 1.80$^{+0.02}_{-0.02}$ & 1.82$^{+0.02}_{-0.02}$ & 
1.79$^{+0.02}_{-0.02}$ \\
$r_\rmn{in}$ & 219$^{+81p}_{-119}$ & 24$^{+12}_{-12}$ & 300$^{+p}_{-97}$ & 
74$^{+114}_{-35}$ & 136$^{+164p}_{-77}$ & 103$^{+197p}_{-51}$ 
& 92$^{+147}_{-45}$ \\
$\log\xi$ & 2.33$^{+0.37}_{-0.14}$ & 2.30$^{+0.30}_{-0.24}$ & 2.70$^{+0.05}_{-0.31}$ &
2.51$^{+0.25}_{-0.32}$ & 2.72$^{+0.18}_{-0.42}$ & 2.30$^{+0.09}_{-0.19}$ & 
2.45$^{+0.33}_{-0.35}$ \\
$\mathcal{R}$ & 0.32$^{+0.11}_{-0.11}$ & 0.56$^{+0.14}_{-0.16}$ & 0.25$^{+0.07}_{-0.08}$ &
0.21$^{+0.13}_{-0.05}$ & 0.25$^{+0.06}_{-0.07}$ & 0.33$^{+0.06}_{-0.09}$ & 
0.20$^{+0.12}_{-0.06}$ \\
\hline 
\multicolumn{8}{c}{\texttt{relxilllp (lamp-post)}} \\
$\Gamma$ & 1.84$^{+0.02}_{-0.02}$ & 1.99$^{+0.03}_{-0.03}$ & 1.66$^{+0.02}_{-0.02}$ &
1.85$^{+0.02}_{-0.02}$ & 1.80$^{+0.02}_{-0.02}$ & 1.82$^{+0.02}_{-0.02}$ & 
1.78$^{+0.03}_{-0.01}$ \\
$h$ & 164$^{+136p}_{-112}$ & 25$^{+32}_{-14}$ & 202$^{+98p}_{-135}$ &
36$^{+264p}_{-22}$ & 109$^{+191p}_{-80}$ & 118$^{+132}_{-82}$ & 
43$^{+257p}_{-28}$ \\
$r_\rmn{in}$ & 110$^{+45}_{-104p}$ & 7$^{+18}_{-1p}$ & 145$^{+64}_{-139p}$ &
69$^{+73}_{-63p}$ & 110$^{+52}_{-104p}$ & 68$^{+42}_{-62p}$ & 
86$^{+91}_{-80p}$ \\
$\log\xi$ & 2.43$^{+0.28}_{-0.16}$ & 2.35$^{+0.36}_{-0.20}$ & 2.69$^{+0.08}_{-0.30}$ &
2.54$^{+0.22}_{-0.35}$ & 2.71$^{+0.18}_{-0.41}$ & 2.30$^{+0.10}_{-0.19}$ & 
2.48$^{+0.30}_{-0.37}$ \\
\hline
$h_\rmn{isco}$ & 256$^{+44p}_{-101}$ & 26$^{+32}_{-14}$ & 300$^{+p}_{-70}$ &
$239^{+61p}_{-90}$ & 272$^{+28p}_{-73}$ & 175$^{+91}_{-47}$ & 
$300^{+p}_{-119}$ \\
\hline
\end{tabular}
\flushleft
\small{\textit{Notes.} While $q = 3$ in the general case, in the lamp-post configuration 
the disk emissivity profile and the reflection fraction ($\mathcal{R}$) are self-consistently 
determined by the height of the source ($h$, in gravitational radii). The ionization parameter 
($\xi$) is in units of \xicgs. In both models $a^* = 0$ and $i = 30$\degr are assumed. Also 
reported is the height value for a disk extending down to the innermost stable circular orbit 
($h_\rmn{isco}$).}
\end{table}

%%%%%%%%%%%%%%%%%%%%%%%%%%%%%%%%%%%%%%%%%%%%%%%%%%%%%%%%%%%%%%%%%
\section{\fek emission variability}

\subsection{Long timescales (months to years)}

\begin{figure}
\centering
\includegraphics[width=14cm]{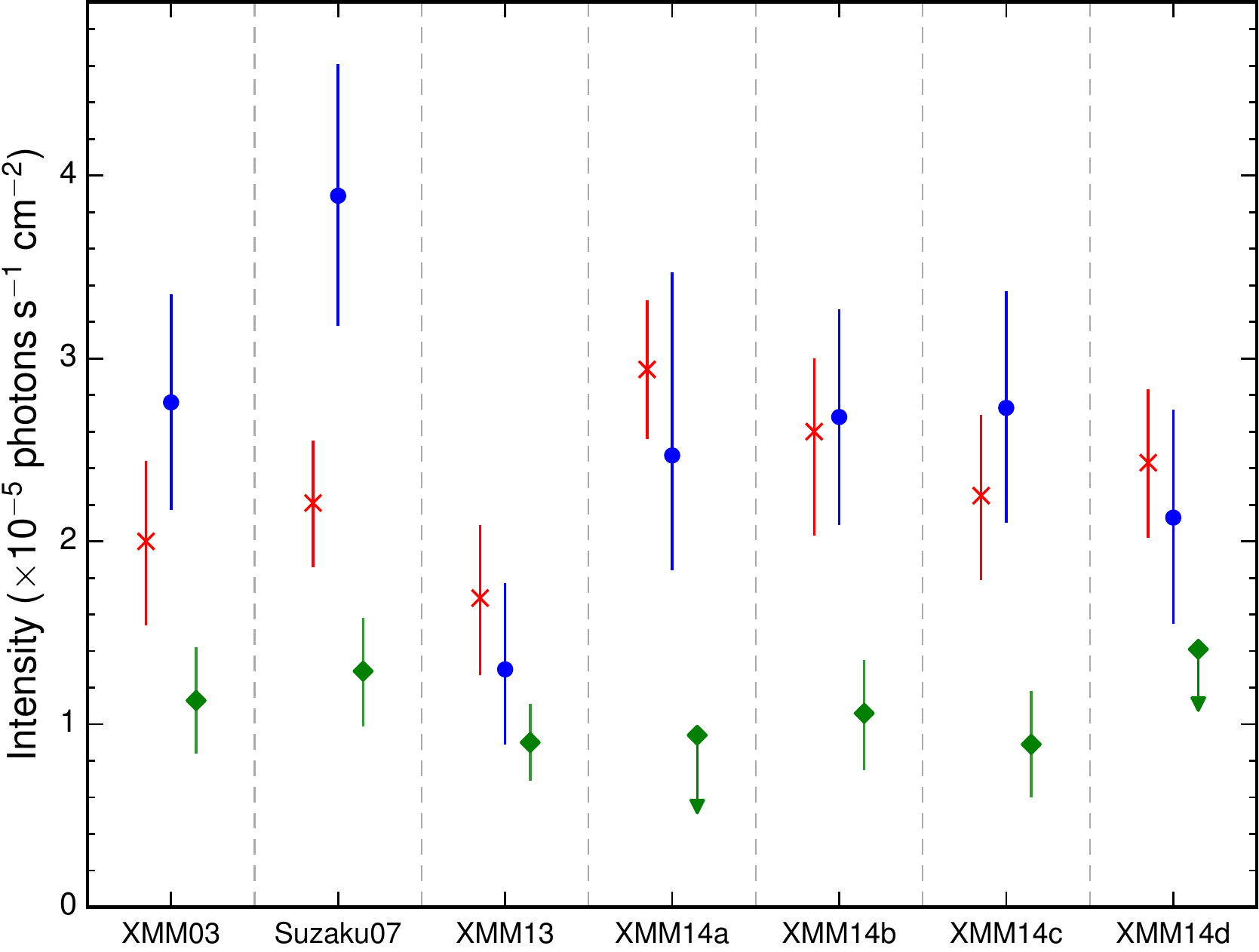}
\caption{Intensity of the three \feka emission features in the 2003, 2013, 2014 
\xmm, and 2007 \suzaku spectra of \ark: narrow \fei (6.40 keV, red crosses), 
tentative \fexxvi (6.98 keV, green diamonds), and disk line ($\sim$\,6.5 keV, blue 
dots). While the two narrow lines have a nearly constant flux, the broad component 
exhibits a marked drop in 2013, which is significant at the 5$\sigma$ and 2.5$\sigma$ 
levels compared to 2007 and 2014a, respectively (the plotted error bars correspond 
to the 90\% confidence intervals). For simplicity, a cross calibration of 1.00 
is assumed between \xmm/pn and \suzaku/XIS. This should be correct within 10\% 
(Tsujimoto et al. 2011).} 
\label{vs}
\end{figure}

To obtain a more quantitative evaluation of the variability of the intensity of all the \fek 
lines across the various epochs, we reverted back to the \relline model and fitted the seven 
spectra at 3--10 keV with the usual assumptions on spin, inclination, and emissivity index for 
the disk line ($\chidof = 1650/1575$; see footnote 3). The results are plotted in Figure~\ref{vs}, 
and are largely model-independent, since a similar trend would be found by fitting the broad 
component as a Gaussian ($\dchi \simeq 11$). The strength of the supposed BLR lines at $E_1 = 
6.399 \pm 0.006$ keV and $E_2 = 6.979 \pm 0.014$ presents little in the way of fluctuations. 
The latter is consistent with being constant, while the former could be somewhat more variable, 
although this might be due to the superposition with the disk line, whose exact profile is 
unknown and hard to disentangle. The relativistic feature, in fact, is definitely erratic. In 
the 2013 low-flux state, the disk line drops by a factor of two compared to the closest 
observations, with a significance of 3.5$\sigma$ with respect to the average 2014$m$ spectrum 
(and 2--3$\sigma$ over the individual sequences). The difference is even more pronounced (a 
factor of three at 5$\sigma$) considering the \suzaku spectrum, which, however, was taken six 
years earlier. 

\begin{figure}
\centering
\includegraphics[width=14cm]{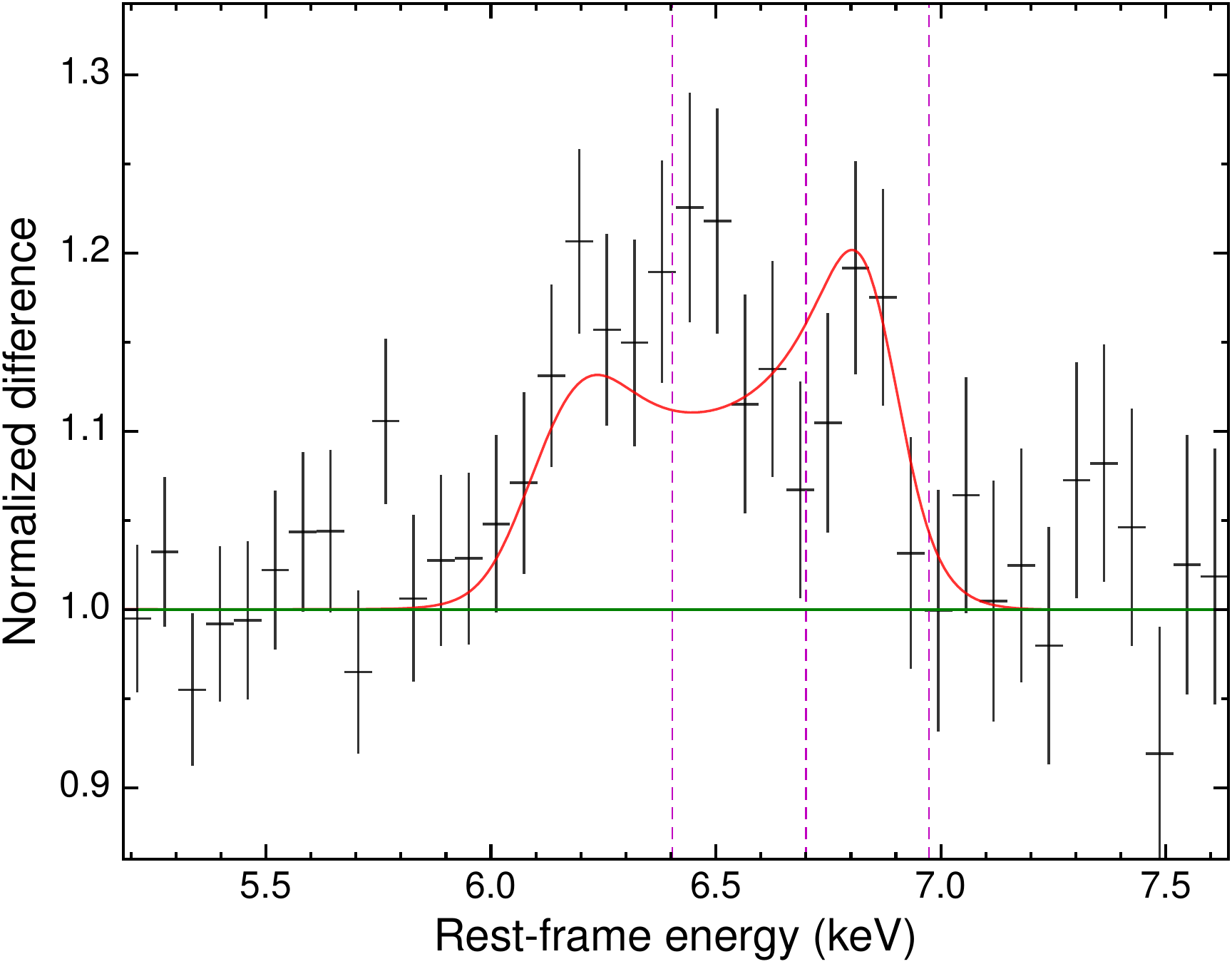}
\caption{Zoom in on the \fek emission band of the `2014--2013' difference spectrum,
normalized to take into account the different slopes of the hard X-ray continuum in
the two epochs and rebinned for clarity. The correction for the relative energy
calibration issues has also been applied. It stands out how the emission-line complex
is more prominent in the high-flux state. The variations mainly involve the broad
component, although an excess is also seen in proximity of the narrow 6.4-keV core.
For comparison, the red curve shows the profile of a disk line with energy of 6.6 keV
emitted between 60 and $120\,\rg$ (for  $a^* = 0$, $i = 30$\degr, and $q = 3$; note 
that this is not a fit). The \fei, \fexxv, and \fexxvi \ka positions are traced for 
guidance.}
\label{ds}
\end{figure}

Since the disk line seems to vary not only in flux but also in shape, and specifically in the 
extent of the red wing, we further investigated its evolution with time by computing the difference 
spectrum between the 2014/high and 2013/low states. In order to maximise the signal-to-noise, 
given that the calibration of the relative energy scales might still be not perfectly matched 
after the gain correction (Appendix~A), we used the merged 2014$m$ spectrum. Once the effects of 
the different hard X-ray continuum slopes are factored out, the outcome is the one illustrated in 
Figure~\ref{ds}. The \fek emission complex clearly underwent a substantial change from 2013 to 
2014. The variable component has an uneven shape, vaguely reminiscent of a double peak, where a 
rising wing at 6 keV can be easily recognized. Notably, no discrepancy is revealed at the \fexxvi 
\ka energy, while the larger excess is visible around 6.4 keV. We can then speculate that part of 
the neutral-iron fluorescence does not occur at BLR scales, but closer in (possibly on the outskirts 
of the disk), thus responding to the continuum on shorter timescales. Even in this model-independent 
perspective, however, the broad emission line unveiled in the spectral analysis undoubtedly plays 
a major role in the overall \fek variability. For comparison, in Figure~\ref{ds} we also show the 
profile of a disk line of intensity $2 \times 10^{-5}$ photons s$^{-1}$ cm$^{-2}$, centered at 6.6 
keV, and arising from an annulus between 60 and $120\,\rg$ (see Table~\ref{tx} and Figure~\ref{vs}). 
Aside from around 6.4--6.5 keV, the agreement with the variable component is excellent. 

In this light, it is plausible that different regions of the disk contribute to the \fek emission 
at different times. This would explain the disparity between the values of the disk inner radius 
returned by the single spectra, and the moderate shifts of the line's energy could be ascribed to 
the local ionization state of the gas. Once the eventualities of a disk truncation and/or of a 
high-elevation source are regarded as unlikely (Section 3.4), the broad \fek line properties imply 
that, for any reason (e.g., full iron ionization, extreme velocity blurring, Comptonization), the 
inner disk does not produce a detectable \ka feature. In this view, these regions could yield a 
featureless \textit{pseudo} continuum, while a number of hotspots associated with localized magnetic 
flares further out (e.g., Galeev et al. 1979; Dov{\v c}iak et al. 2004a; Goosmann et al. 2007) 
might dominate the \fek emission. Indeed, at intermediate (i.e., tens of $\rg$) radii, the 
illumination from a standard (centrally concentrated, compact, and low-height) X-ray corona would 
be less effective, and an additional source would then be required to sustain the local emissivity. 
Following on, we modified the \relline model assuming an annulus of width 10\,$\rg$ with gray 
($q = 0$) emissivity as the most relevant reflecting surface to the formation of the observed 
disk line. As could be expected, the fit of the seven spectra is virtually as good as in the 
original case ($\chidof = 1652/1575$), with the inner border of each ring commensurate with the 
usual values for $r_\rmn{in}$ (Table~\ref{tx}). Emission from the entire annulus is only possible 
if there are at least a few active flares/hotspots at a time. At 60\,$\rg$, in fact, any structure 
in Keplerian motion only covers a small fraction ($\Delta \phi \sim 20\degr$) of its orbit in the 
120-ks span of a single \xmm observation. In this respect, we switched to the more flexible 
\texttt{kyr1line} profile (Dov{\v c}iak et al. 2004b), which also enables the selection of annular 
sectors. By freezing either the inner radius or the line's rest energy from the previous fit to 
control the degeneracy with the central azimuth, we confirmed that the opening angle of each 
sector is always consistent with being $\Delta \phi \ga 180\degr$, since approaching as well as 
receding parts of the disk must be involved to account for both the red and the blue wing of the 
\fek line. 

\subsection{Short timescales (days to hours)}

Depending on the number, radial distance, and lifetime of the putative hotspots, intensity and 
shape variations in the \fek emission complex of \ark can occur even on timescales much shorter 
than one year. In order to appraise any rapid change eluding the analysis of the time-averaged 
spectra, we applied to the four 2014 \xmm sequences an `excess map' technique, which has proven 
to be very effective in revealing the possible modulation of redshifted \fek features in some 
Seyferts (Iwasawa et al. 2004; Tombesi et al. 2007). We outline here the main steps of this 
procedure, referring the reader to the previous works for a more detailed description (see also 
De Marco et al. 2009). We adopted a resolution of 100 eV in energy and 5 ks in time. Since the 
orbital period at the innermost stable circular orbit of a Schwarzschild black hole with mass of 
$1.5 \times 10^{8}\,M_{\sun}$ is about 70 ks (e.g., Bardeen et al. 1972), the latter choice ensures 
that the reasonable variability timescale in the `rotating hotspot' scenario is largely oversampled. 
We then considered the first 115 ks of each observation, neglecting less than 11 ks of the 
cumulative good-time interval in the rounding. Given the reduced live time of the EPIC/pn Small 
Window mode, the net exposure of the single $23 \times 4$ slices is 3.5 ks. The resulting 
time-resolved spectra were grouped to a 5$\sigma$ significance per channel, and fitted with a 
simple power law in the 3--5 plus 7--10 keV range. We worked in the observed frame, and did not 
correct the energy scale for the calibration problems. In fact, these are not straightforward to 
tackle, and should affect all the data sets in the same way. The energies of the two narrow lines 
(6.40 and 6.98 keV at rest) are then shifted to 6.26 and 6.82 keV.

\begin{figure}
\centering
\includegraphics[width=9cm]{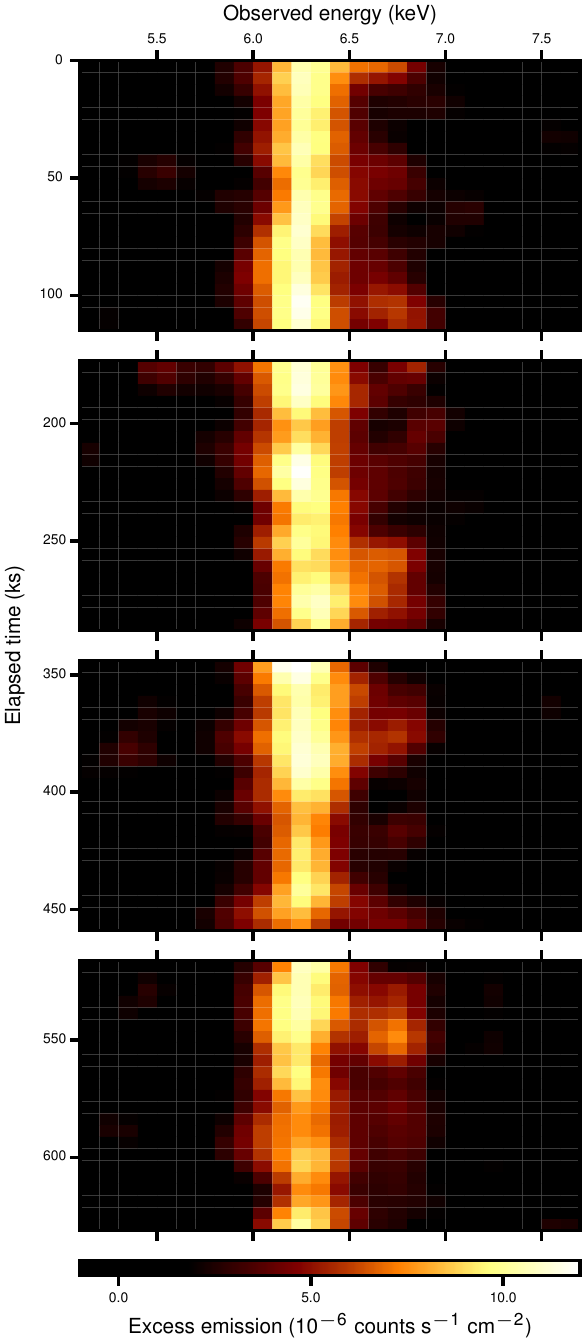}
\caption{Excess emission map in the \fek band for the 2014 \xmm observation, with a resolution 
in the energy--time plane of 100 eV $\times$ 5 ks. The image has been smoothed with an elliptical 
Gaussian kernel with FWHM of 250 eV $\times$ 15 ks. Energies are in the observed frame, and no 
gain correction was applied. The two narrow lines from the spectral models are thus centered at 
6.26 and 6.82 keV. The lower (negative) value of the color scale indicates the 1$\sigma$ noise 
level, so the main features in the map are expected to be highly significant. As a gauge, 
considering the effective exposure, an excess of $5 \times 10^{-6}$ photons s$^{-1}$ cm$^{-2}$ 
corresponds to $\sim$\,12 counts. (Note that the time gap between the sequences is not to scale).} 
\label{em}
\end{figure}

Within the four sequences, a constant trend of the photon index yields a $\chi^2$ between 21 and 
25 for $\nu = 22$ (the typical 90\% uncertainty on the single estimates of $\Gamma$ is $\pm0.07$). 
The average values and 1$\sigma$ dispersions are $\Gamma = 1.810\,(\pm 0.043)$, $\Gamma = 
1.762\,(\pm 0.041)$, $\Gamma = 1.771\,(\pm 0.040)$, $\Gamma = 1.744\,(\pm 0.045)$, respectively. 
Once the continuum is determined, the excess emission map is computed from the residuals in the 
100-eV binned spectra, corrected for the detector's collecting area. To verify that all the pixels 
in the energy--time plane have enough statistics, we recorded the number of counts in the 6.0--6.1 
keV bin of each spectral slice. This ranges from 89 to 141, with a mean (median) value of 117 (118). 
As mentioned, any intensity variation is expected to take place on longer timescales than the 
sampling time. It is thus possible to suppress the noise between contiguous pixels and emphasize 
the significant features by applying a low-pass filter (see Iwasawa et al. 2004). We smoothed the 
image through an elliptical Gaussian kernel with $(\sigma_E,\,\sigma_t)$ = (1.06,\,1.27) pixels, 
corresponding to a FWHM of 250 eV $\times$ 15 ks (i.e., the light-crossing time of $20\,\rg$). 
The final excess emission map is shown in Figure~\ref{em}, and clearly demonstrates that the \fek 
line complex is extremely dynamic. The track of the main, 6.4-keV narrow line is the most 
persistent, yet it is characterized by intensity fluctuations, especially in the second half of 
the observation. Highly variable structures are detected both redwards and bluewards, with no 
obvious periodicity nor correlation with each other. 

The average noise in the smoothed excess map is $\sim$\,$1 \times 10^{-6}$ photons s$^{-1}$ 
cm$^{-2}$ at the 1$\sigma$ level. All of the hottest blob- and plume-like features in the 
image contain at least a hundred counts, so their transient nature is expected to be genuine. 
Quite surprisingly, also the 7-keV narrow line (whose plain, yet mistaken identification from 
the mere time-averaged spectral analysis would be as \fexxvi \ka from the BLR) erratically brightens 
and fades on timescales of $\sim$\,30--50 ks, and seems to be one and the same with the blue horn 
of the disk line. The absence of persistent, prominent \fexxvi emission has the further advantage 
of alleviating the apparent discrepancy between the \xmm spectra and the \chandra one, where such 
a feature is not evident. In this respect, it is worth noting that the actual overlap between the 
three \chandra/HETG and the four \xmm/EPIC observations is less than 60 ks (Table~\ref{to}), during 
which there was only a moderate flaring activity in the blue \fek excess based on Figure~\ref{em}. 
Some BLR-like emission from \fexxvi cannot be firmly excluded (see Appendix~C), but this is 
impossible to separate from the variable, disk component (and any \fei \kb) with the present data. 

\begin{figure}
\centering
\includegraphics[width=14cm]{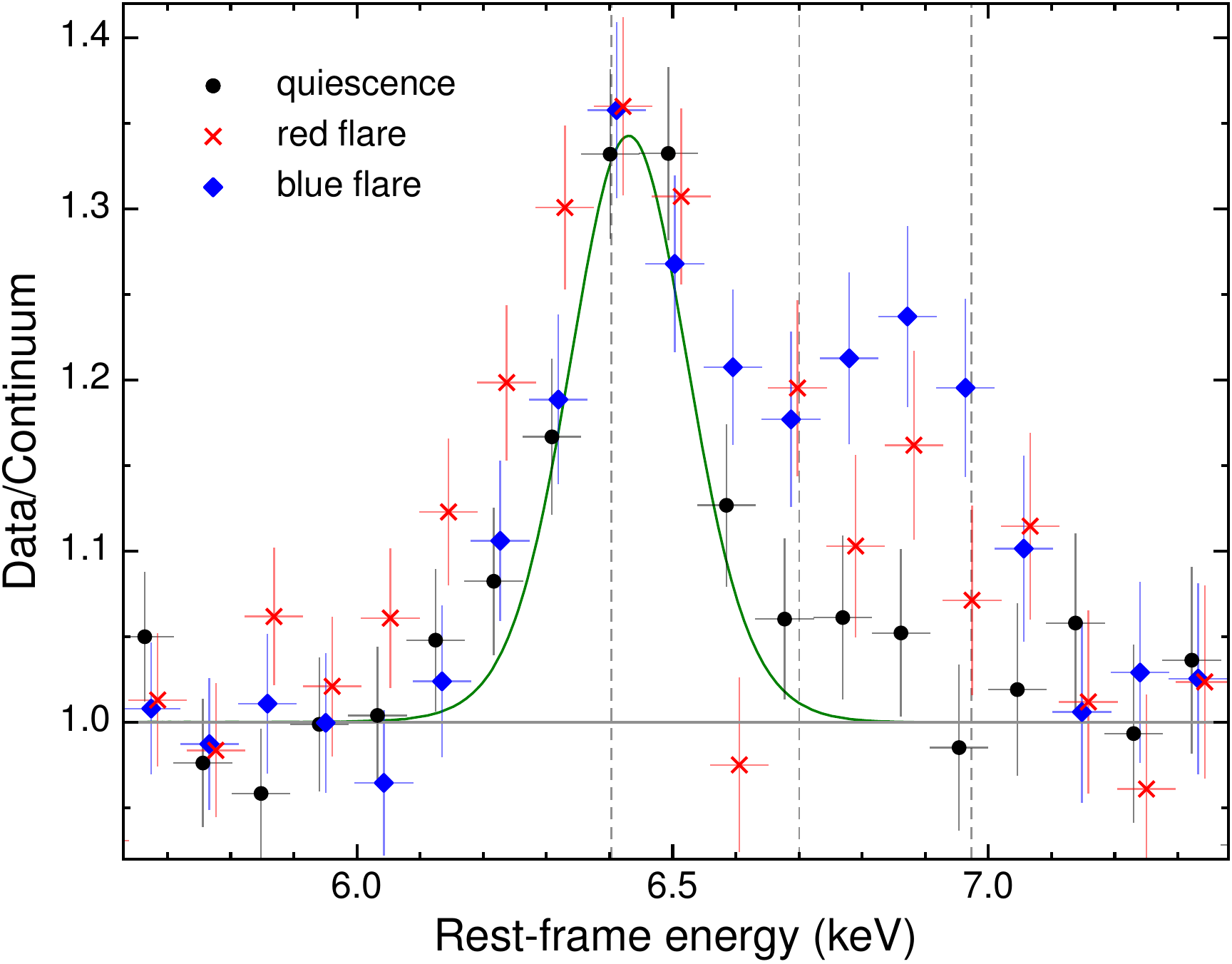}
\caption{Different profiles of the \fek emission complex in three 30-ks intervals selected after 
a visual inspection of the excess emission map in Figure~\ref{em}, and corresponding to quiescent 
(dots; 10--40 ks), red flare (crosses; 75--105 ks), and blue flare (diamonds; 253--283 ks) states. 
Significant changes on short timescales are seen on both sides of the narrow 6.4-keV line (drawn 
in green), conclusively locating the origin of the broad component(s) from the inner accretion 
disk. The data are binned by a further factor of three over the adopted 30-eV resolution for 
visual purposes.} 
\label{hs}
\end{figure}

Based on a qualitative assessment of Figure~\ref{em}, we selected three intervals with duration 
of 30 ks, and extracted the relative spectra (with net exposure of 21 ks each) as representative 
of a \textit{quiescent} \fek state (between 10 and 40 ks of the elapsed time), a \textit{red flare} 
(75--105 ks), and a \textit{blue flare} (253--283 ks), respectively. The resulting \fek profiles 
are plotted in Figure~\ref{hs} as ratios over the power-law continuum, for which a common photon 
index was assumed ($\Delta \Gamma < 0.03$ otherwise). During the low-activity phase, only the 
narrow ($\sigma = 40$ eV) Gaussian line at 6.4 keV is statistically required, with just a hint 
of asymmetry on the blue side. An extended red wing emerges later on in the first sequence, 
immediately following and preceding two minor flares in the 6.5--7 keV rest energy band. This 
blue excess is strongest towards the end of the second observation, when the red wing is back at 
its faintest, confirming the lack of simultaneity between the two. We note that similar episodes 
are seen in the last three days of the campaign as well; the spectra are somewhat noisier because 
the parent events are short-lived and/or less intense, but are always found to be in agreement with 
those of Figure~\ref{hs}. 

To quantify the extent of the intensity variations for the red and the blue excess, we modeled the 
three spectra including two additional Gaussian lines besides the main 6.4-keV one. This is just a 
rough approximation, but the current, time-resolved data are not very sensitive to much finer profiles 
(see below). We fitted a local power law at 4--9 keV to avoid the subtle continuum complexities (e.g., 
Figure~\ref{re}) as well as bins with $<$\,5$\sigma$ significance at the higher end (a 30-eV resolution 
is still adopted). The photon index and the amplitude of the central feature were tied between the 
three states, since they do not appear to vary on statistical grounds ($\dchidnu = -2/-$4 when 
both are free). The joint fit is very good indeed, with $\chidof = 500/485$. The red and the blue 
line have rest energies of 6.19 and 6.83 keV, and widths of $\sim$\,60 and 125 eV, respectively 
(Table~\ref{tk}). The intensity of the red \fek excess increases over the quiescent state and 
subsequently drops during the blue flare with a 1.7$\sigma$ and 2.1$\sigma$ confidence (see also 
Appendix~D), while the enhancement of the blue excess is significant at the 3.7$\sigma$ level. If 
the observed variations are due to the differential beaming effects along the orbit of a rotating 
hotspot on the disk surface, and about one fourth of a Keplerian period ($T_\rmn{K}$) is needed for 
them to be detectable, the characteristic 30--50 ks timescale implies a typical distance from the 
central black hole of $\sim$\,10\,$\rg$. Although the large SMBH mass in \ark and the gaps between 
the sequences do not facilitate a visual recognition, there does not seem to be any regular modulation 
of the red and blue features, in either strength or energy. This suggests that the hotspots might 
not survive for an entire orbit ($T_\rmn{K} \simeq 200$ ks at 10\,$\rg$, and 650 ks at 27\,$\rg$), 
but they should be continuously regenerated. In this scenario, the same red wing and the source 
of the possible changes seen in the 6.4-keV core (Figure~\ref{ds}) could be explained as the blue 
horns of \fek lines arising closer in in the disk. 

%%%%%%%%%%%%%%%%%%%%%%%%%%%%%%%%%%%%%%%%%%%%%%%%%%%%%%%%%%%%%%%%%
\section{Discussion}

\ark is an X-ray bright local Seyfert galaxy, prototype of the \textit{bare} AGN subclass, which 
is characterized by a clean line of sight to the central source. This allows an unbiased study of 
some critical components that are commonly found among active galaxies, such as the soft and/or 
hard X-ray excess, and the possible relativistically blurred iron line. Due to its uniqueness, 
\ark was simultaneously targeted in a concerted 2014 campaign by \xmm, \chandra/HETG, and \nustar. 
In this work we have taken advantage of the combination of high energy resolution and high effective 
area afforded by these new data sets in the 6--7 keV band, and revisited the high-quality \xmm 
and \suzaku archival observations to understand the nature, properties, and variability of the 
distinctive \fek emission features. 

The interpretation of the \fek emission complex in \ark is definitely challenging. The same 
intricate shape implies a blending of different components that are not all readily identifiable, 
with the only exception of the neutral \feka at 6.4 keV. Moreover, the time-averaged spectral 
analysis can be rather misleading, if not validated by the supplementary variability study. The 
presence of a constant, narrow line from \fexxvi at 6.97 keV, for instance, seems to be a mere 
coincidence, deriving from the combination of some possible iron emission from the BLR (H-like 
\ka and/or neutral \kb) and the blue part of a rapidly variable feature from the disk, when the 
latter is integrated over sufficiently long ($>$\,100 ks) exposures. In this sense, the excess 
emission map reveals that the broad line itself is likely the random superposition of narrower 
features from a few, disconnected regions on the disk surface, as opposed to a single circle or 
annulus. Taken individually, the presumable constituents of the erratic \fek fluorescent component 
look neither unresolved nor heavily smeared, and are not subject to large displacements from their 
usual rest energy (6.4--7 keV). Consequently, they cannot be plainly classified under any of the 
two flavors of \fek emission transients that have been found among AGNs, namely narrow and broad, 
which typically exhibit substantial redshifts. While the sporadic detection of narrow lines at 
4--5.5 keV (e.g., Turner et al. 2002; Porquet et al. 2004; Petrucci et al. 2007) might be 
controversial (see Vaughan \& Uttley 2008 for a critical standpoint), the broad \fek transients 
are more intriguing, in particular when the evolution with time of the line's flux and/or centroid 
can be trailed (e.g., Iwasawa et al. 2004; Turner et al. 2006). 

The hotspot picture (e.g., Nayakshin \& Kazanas 2001; Dov{\v c}iak et al. 2004a) is widely invoked 
for both manifestations of variable \fek features, since it naturally delivers narrow line profiles 
as well as sizeable intensity and energy changes thanks to the compact emitting region and its 
orbital motion in a relativistic gravitational field. This is a promising explanation also for 
the \fek behavior observed in \ark, where the relative contribution of different radial and/or 
azimuthal segments of the disk surface could drive the variability of the reprocessed \fek emission 
on timescales of several hours to a few days. To test the physical applicability of the hotspot 
model to \ark, we fitted once again the quiescent and red/blue flare spectra replacing the two 
Gaussians for the red and blue \fek excess with \texttt{kyr1line} components, originating from 
annular sectors with radial width of $10\,\rg$ and flat emissivity ($q = 0$, plus $a^* = 0$, 
$i = 30$\degr). We assumed that the rest energy of both lines is 6.97 keV and, for simplicity, 
that the same regions of the disk are active in the flaring states. This is unrealistic, of course, 
but the combinations of line energy (i.e., ionization), distance, and azimuth that can give rise to 
the detected features are manifold, and it is not feasible to discriminate between them all. A 
self-consistent solution was ensured by linking the angular width of each annular sector (in 
degrees) to its inner radius through the expression $\Delta \phi = 360\degr \Delta t/T_\rmn{K}(r) 
= 1.16\,M_7^{-1}(r_\rmn{in} + 0.5\Delta r)^{-1.5}\Delta t$ (adapted from Bardeen et al. 1972), 
where $M_7 = 15$ is the black hole mass in units of $10^{7}\,M_{\sun}$, $\Delta r = 10\,\rg$, 
and $\Delta t = 30$ ks. While a smaller $\Delta r$ would be probably more appropriate for an 
isolated hotspot, flares are thought to come in avalanches (Poutanen \& Fabian 1999), hence a 
wider area might be coherently illuminated. We also note that larger structures would be disrupted 
more quickly by differential rotation, in agreement with the short-lived events observed here. 

\begin{figure}
\centering
\includegraphics[width=14cm]{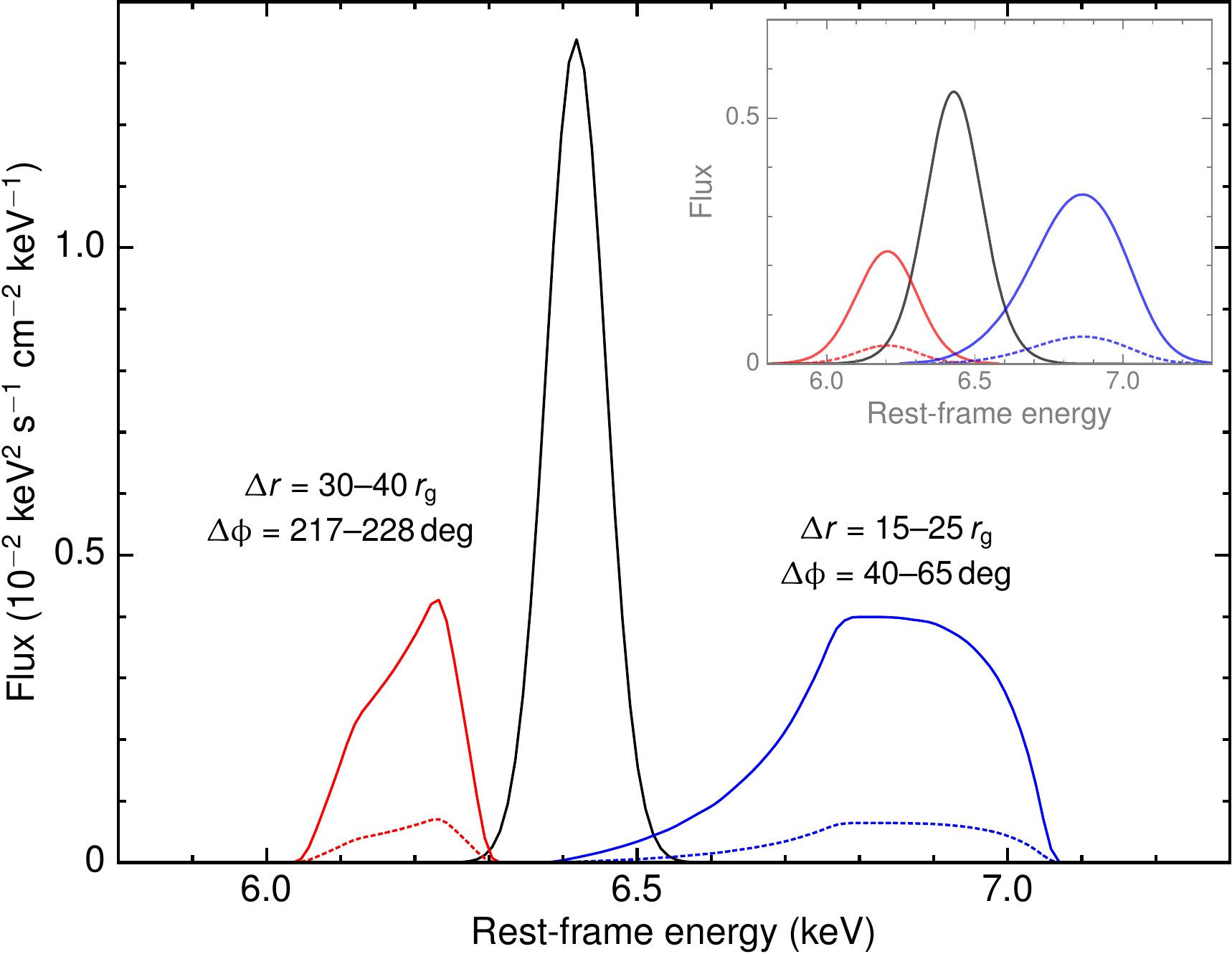}
\caption{\fek emission components in the quiescent, red flare, and blue flare spectral 
states for the hotspot model. The narrow, BLR-like \ka from neutral iron at 6.4 keV (in 
black) is constant, while the highest (solid) and lowest (dashed) intensities are shown 
for the red and the blue transient features. The radial and azimuthal coordinates of the 
relative active regions are also printed, for $a^* = 0$, $i = 30$\degr, and a rest energy 
of the line of 6.97 keV ($\phi = 0\degr$ corresponds to the maximal approaching speed).
The profiles are computed assuming an ideal response with linear resolution of 10 eV, 
while the inset shows their smearing after convolution with the EPIC/pn response (note 
that the vertical scale is compressed by a factor of two). The intrinsic asymmetry is 
almost completely lost at CCD resolution.} 
\label{ky}
\end{figure}

The resulting fit statistic is exactly the same of the phenomenological attempt above ($\chidof = 
500/485$). Indeed, as the range of angles involved is limited ($\Delta \phi = 11\degr$ for the 
red horn and 25\degr for the blue one), the emitted lines are narrow, and lose any asymmetry after 
the convolution with the detector's response, becoming in practice indistinguishable from simple 
Gaussian profiles (Figure~\ref{ky}). We chose a framework where positive angular velocity 
corresponds to counterclockwise rotation, and $\phi = 0\degr$ to the maximal Doppler blueshift 
for matter moving towards the observer. The red feature then stems from a sector delimited by 
$r \simeq 30$--$40\,\rg$ and $\phi \simeq 217$--228\degr, while the boundaries for the blue one 
are $r \simeq 15$--$25\,\rg$ and $\phi \simeq 40$--65\degr. The significance of their variability 
is confirmed at the 2.2$\sigma$ and 3.6$\sigma$ levels. Once the uncertainties on the best-fit 
parameters are considered (Table~\ref{tk}), it is explicit that the two active regions above, 
responsible for the red (6.0--6.3 keV) and the blue (6.5--7.0 keV) \fek excess, are widely 
separated and respectively confined to the receding and approaching parts of the disk. Although 
there is some overlap in the range of distances they are allowed to lie at, between 
$\sim$\,20--$30\,\rg$, in the absence of clear periodic patterns in the excess emission map a 
single hotspot appears unlikely. On the other hand, some arc-like structures (like at $t \sim 
200$--230 ks; Figure~\ref{em}) are strongly suggestive of orbital motions. Trying to describe 
the whole variability of the \fek features in these terms, however, goes beyond the scope of 
the present work.

An energy reservoir formed by an ensemble of buoyant magnetic loops/blobs has been recurrently 
proposed to explain the X-ray spectra and light curves of AGNs (e.g., Galeev et al. 1979; Haardt 
et al. 1994; Merloni \& Fabian 2001; Czerny et al. 2004). The stored energy would be released 
through field reconnection events, which sustain the energetic stability of the corona and, in 
principle, can induce appreciable flares in the primary X-ray emission. Part of this radiation 
impinges on the disk below, and is reflected off the surface. Based on the X-ray fractional 
variability and the log-normal flux distribution of AGNs, the number $\mathcal{N}_\rmn{flare}$ 
of uncorrelated active flare/hotspot regions present at any given time must be limited (Uttley 
et al. 2005). In \ark, the amplitude of the intensity fluctuations and the same occurrence of 
quasi-quiescent \fek states are consistent with $\mathcal{N}_\rmn{flare}$ being no more than a 
few. It is not surprising that this coronal component is not the main source of primary X-ray 
emission, then. We can guess the intrinsic, cumulative luminosity of the hotspots by extrapolating 
the reflection component from the \relxill model of Section~3.4. Over the 0.1--100 keV band, this 
is 5--$6 \times 10^{43}$ \lumcgs in the four sequences of the 2014 campaign. For an average 
reflection fraction $\mathcal{R} \simeq 0.25$ (Table~\ref{tr}), we can argue that about one fourth 
of the power-law continuum emerges from the flares, bringing a further 8--$9 \times 10^{43}$ \lumcgs. In the most favorable conditions of zero time lag (i.e., $h \ll r$) and isotropic 
emission (which is not so obvious in the local frame), with just $\mathcal{N}_\rmn{flare} \sim 
3$--5 the single flare/hotspot would thus contribute to $\sim$\,5\% only of the total 0.1--100 
keV luminosity of \ark ($6.5 \times 10^{44}$ \lumcgs; Porquet et al. 2016), and still less 
than 10\% restricting to the 2--10 keV band. The fact that no sudden spike is seen in the light 
curves is therefore plausible. For the same reasons, even the strongest fluorescent lines in the 
soft X-rays (like \oviii \lya at 0.654 keV) should be completely missed in the EPIC/pn spectra, 
while the low RGS effective area prevents any attempt of time-resolved analysis. Interestingly, 
in the model by Haardt et al. (1994) the most luminous coronal blobs are located at $r = 24\,\rg$, 
and for the black hole mass and Eddington rate of \ark (roughly 0.1) they should emit a few 
$\times$10$^{43}$ \lumcgs.

In their recent study on the geometry of AGN coronae, Wilkins \& Gallo (2015) have suggested 
that the very detection of X-ray reflection signatures from the surroundings of the innermost 
stable circular orbit requires a patchy coronal structure. While a number of clumps could appear 
as a continuous, extended medium in long exposures, the reflected spectrum would still be exempt 
from heavy Comptonization effects by leaking through the voids. Our analysis of the \fek transients 
in \ark hints at a hybrid situation, where the X-ray corona is smooth in the center and patchy 
towards the edges. This would simultaneously justify the apparent lack of reflection from the 
inner disk (except, perhaps, for the 2007 \suzaku observation) and the short-term \fek 
variability. It might be appealing to associate the characteristic radius of the transitional 
region where the observed features arise with a change of the physical conditions within the 
accretion flow. A popular Comptonization model like \optxagn (Done et al. 2012), for instance, 
adopts the size of the corona as a key dividing line between the outer and the inner disk, setting 
the energy balance between thermal emission, soft X-ray excess and hard X-ray power law. Matt et 
al. (2014) made use of \optxagn in their broadband fits to the 2013 \xmm/\nustar spectra of \ark, 
and obtained a coronal extent of $\sim$\,10--$30\,\rg$, depending on the chosen black hole spin. 
While remarkably similar to the inferred location of the hotspots, the latter can still be seen 
as a convenient parameterization. However, the existence of a critical distance scale is also 
envisaged on theoretical grounds. Indeed, for Eddington rate values as the one of \ark a 
standard geometrically thin, optically thick accretion disk (Shakura \& Sunyaev 1973) should 
invariably undergo a transition from gas to radiation pressure dominated below a certain radius, 
unless all of the accretion power is dissipated in the corona (Svensson \& Zdziarski 1994). 
Structure and properties of the X-ray source could then be radically different through this 
physical boundary, with the fragmentation of the corona into filaments/clumps at its outskirts. 

Alternatively, the cause of \fek variability can be fully inherent to the disk. In fact, the 
radiation pressure dominated zone was soon realized to be unstable against thermal and viscous 
perturbations (Lightman \& Eardley 1974; Shakura \& Sunyaev 1976). Other dynamical instabilities 
follow in magnetized disks, leading to widespread turbulence (see Balbus \& Hawley 1998 for a 
comprehensive review) and large density gradients (e.g., Begelman 2001). The ensuing inhomogeneity 
significantly alters the spectral features reprocessed from the disk with respect to the usual 
case of a flat, uniform slab. The reflection component would bear the different signs of more 
rarefied versus compact layers, and even moderate density variations can deeply affect the 
intensity of the \fek emission without any correlation with the illuminating continuum 
(Ballantyne et al. 2004, 2005). The continual succession of clumps and voids at any given 
location and the related emissivity changes are then a potential explanation for the \fek 
variability. Relativistic beaming induced by turbulence can also be important (Armitage \& 
Reynolds 2003). Among the various clumping instabilities, of particular relevance to \ark 
could be the so-called `photon bubbles' (Gammie 1998), which, differently from most of the 
other perturbations, can develop on timescales much shorter than the orbital period (Blaes 
\& Socrates 2003). 

A theoretical effort to predict the empirical differences between orbiting hotspots, coronal 
clumps, and disk instabilities could be worthwhile in the future. Regardless of the actual 
mechanism responsible for the transient \fek emission observed in \ark, however, our findings 
prove that the environment pervaded by the inner accretion flow is almost certainly highly 
dynamic and/or inhomogeneous. If these conditions, and the underlying physical processes, are 
established to be common among AGNs, there would also be several compelling implications on the 
origin of the broad \fek lines detected in many sources. In fact, some of these profiles could 
result from the integrated contribution of individual, disconnected regions, especially at lower 
(10$^6$--10$^7\,M_{\sun}$) black hole masses, hence proportionally shorter orbital timescales. 

\begin{table}
\centering
\caption{\fek excess features in the flare/hotspot scenario.}
\label{tk}
\begin{tabular}{lccc@{\hspace{35pt}}lcc}
\hline \hline
 & \multicolumn{2}{c}{\texttt{zgauss}} & & & \multicolumn{2}{c}{\texttt{kyr1line}} \\
\hline
 & \textit{red} & \textit{blue} & & & \textit{red} & \textit{blue} \\
$E$ & 6.19$^{+0.07}_{-0.07}$ & 6.83$^{+0.05}_{-0.04}$ & & 
$r_\rmn{in}$ & 30$^{+12}_{-14}$ & 15$^{+6}_{-4}$ \\
$\sigma$ & $<$\,169 & 125$^{+57}_{-46}$ & & 
$\phi$ & 217$^{+33}_{-127p}$ & 40$^{+13}_{-13}$ \\
$I_\rmn{f}$ & 1.66$^{+1.05}_{-0.76}$ & 3.01$^{+0.93}_{-0.80}$ & & 
$I_\rmn{f}$ & 1.58$^{+1.15}_{-0.67}$ & 2.97$^{+1.02}_{-0.83}$ \\
$I_\rmn{q}$ & $<$\,1.02 & $<$\,1.27 & & $I_\rmn{q}$ & $<$\,0.98 & $<$\,1.25 \\
\hline 
\end{tabular}
\flushleft
\small{\textit{Notes.} In both the Gaussian and disk line cases, the continuum 
photon index is $\Gamma = 1.83$\,($\pm$0.03), and the energy of the narrow \ka 
is $E = 6.42$\,($\pm$0.02). The intensities of the variable lines during the 
flaring ($I_\rmn{f}$) and quiescent ($I_\rmn{q}$) states are in units of 10$^{-5}$ 
photons s$^{-1}$ cm$^{-2}$. The rest energy of each \texttt{kyr1line} component is 
fixed at 6.97 keV (see the text for the other assumptions of the model). The lower 
azimuth of the emitting sector ($\phi$) is measured in degrees from the position 
of maximal approaching speed. Errors are given at the 90\% confidence level.}
\end{table}

%%%%%%%%%%%%%%%%%%%%%%%%%%%%%%%%%%%%%%%%%%%%%%%%%%%%%%%%%%%%%%%%%
\section{Conclusions}

We have presented the analysis of the composite emission spectrum due to iron K-shell 
fluorescence in the local Seyfert galaxy \ark, the nearest and brightest example of a bare 
AGN, based on a large 2014 campaign consisting of simultaneous \xmm, \chandra/HETG, and \nustar 
observations. Overall, \ark displays a broad and irregular emission-line complex in the 6--7 
keV energy range, where the main feature corresponds to the \ka transition in neutral iron 
around 6.4 keV. Its profile is well resolved by the \chandra grating to a FWHM of $\sim$\,4700 
\kms, consistent with the typical velocity broadening of the optical broad-line region. The 
narrow \fei \ka accounts for less than one half of the total line emission over this band, 
though. The residuals take the shape of a red wing down to $\sim$\,6 keV and of a blue plateau 
across 6.7 keV, connecting to a second, apparent feature centered just below 7 keV. Energy, 
equivalent width, and variability rule out an interpretation of the red excess as the Compton 
shoulder of the main, 6.4-keV \ka line. The wing's strength reached a maximum during a 2007 
\suzaku observation, dropped by a factor of $\sim$\,2--3 following a low-flux state of the 
X-ray source in 2013, and then recovered one year later. This behavior suggests the presence 
of a \fek reflection component from the accretion disk. Indeed, a model including two narrow 
($\sigma \simeq 40$ eV) lines at 6.40 and 6.98 keV plus a mildly distorted relativistic disk 
line successfully reproduces the entire \fek emission complex in \ark over the different epochs. 
From the time-averaged spectra alone the intensity of the former pair would seem approximately 
constant, as might be expected for BLR gas, only sensitive to the average continuum over long 
timescales. The 7-keV narrow feature would thus be (mistakenly) identified with the \fexxvi \ka, 
possibly blended with the neutral \kb. Some additional fluctuations around 6.4--6.5 keV could 
be ascribed to the elusive broad component, which is responsible for most of the short-term 
spectral changes. 

The disk-line parameters point to an origin in a moderate relativistic regime, at a few/several 
tens of gravitational radii of distance from the central black hole. Rather than a disk truncation, 
this likely implies that the main X-ray corona, if extended, smooths out any reflection from the 
inner disk. Diffuse flares associated with magnetic reconnection events might be involved instead 
to efficiently illuminate the line-emitting regions further out. Such a conjecture is supported by 
the \fek excess emission map obtained from the 2014 \xmm monitoring, which covered four consecutive 
satellite orbits for a span of 7.5 days. This shows that both the red wing and the bulk of the 
6.5--7 keV excess (encompassing also the tentative, BLR-like \fexxvi line) undergo significant 
intensity variations in about $\sim$\,30--50 ks (i.e., 10--15 hours), and that they are substantially 
uncorrelated with each other. The broad \fek emission feature detected in \ark could then be the 
superposition of several different peaks arising from short-lived, yet continuously generated 
orbiting hotspots on the accretion disk surface. Perhaps similar observable effects could be 
produced also as a result of clumpiness in the corona and/or density inhomogeneities in the disk, 
provided that these are highly dynamic. 

Future X-ray observatories like \athena (Nandra et al. 2013) will afford the higher energy 
resolution and larger effective area in the 6--7 keV band that are needed to reveal any fine 
structures in the observed \fek profile and to perform a proper time-resolved spectral analysis, 
so disentangling the various BLR and disk contributions and tracing the evolution of the 
putative hotspots. Thanks to its bare nature, \ark stands out as the most promising source 
to study the accretion disk/X-ray corona system in AGNs, and its possible flaring, transient 
components. 

%%%%%%%%%%%%%%%%%%%%%%%%%%%%%%%%%%%%%%%%%%%%%%%%%%%%%%%%%%%%%%%%%
%\section*{Acknowledgments}
\vspace*{0.5cm}

\noindent
The authors would like to thank the anonymous referee for their useful comments that helped 
improving the clarity of this paper. EN is supported by the UK Science and Technology Facilities 
Council under grant ST/M001040/1. DP acknowledges financial support from the French Programme 
National Hautes Energies (PNHE) and the EU 7th Framework Programme FP7 (2013--2017) under grant 
agreement number 312789. JNR acknowledges support from \chandra grant number GO4-15092X and 
NASA grant NNX15AF12G. AL acknowledges support from the UK STFC. The results presented in this 
paper are based on data obtained with the \chandra \textit{X-ray Observatory}; \xmm, an ESA 
science mission with instruments and contributions directly funded by ESA member states and 
NASA; and \suzaku, a collaborative mission between the space agency of Japan (JAXA) and NASA. 
We have made use of software provided by the \chandra \textit{X-ray Center} (CXC) in the 
application package \textsc{ciao}. The figures were generated using \texttt{matplotlib} 
(Hunter 2007), a \textsc{python} library for publication of quality graphics.

%%%%%%%%%%%%%%%%%%%%%%%%%%%%%%%%%%%%%%%%%%%%%%%%%%%%%%%%%%%%%%%%%
\appendix

\section{Gain correction}
Even before embarking on a proper spectral fitting, we noticed that, in each of the four 2014 
\xmm data sets, the stronger among the K-shell emission features is peaked at 6.45--6.47 keV. 
Rather than signifying a mild ionization state of the gas (e.g., Fe\,\textsc{xix}; Kallman et 
al. 2004), this is due instead to a known inaccuracy in the calibration of the EPIC/pn energy 
scale for some of the most recent \xmm observations (as also discussed in Marinucci et al. 
2014), and the actual line identification is with the neutral \ka. Indeed, in the simultaneous 
\chandra data, its centroid falls at $E = 6.416^{+0.016}_{-0.017}$ keV (Section~3.1). Overall, 
this $\sim$\,50-eV shift in the iron-K energy range is not the most conspicuous shortcoming. 
Regardless of the model adopted for the continuum, a clear residual structure is left around 
$\sim$\,2.2--2.3 keV, in correspondence with the gold absorption edge in the detector's effective 
area. While not strictly relevant to the analysis presented in this paper, this feature was very 
useful to mitigate the calibration problems. As a first approximation, we used the \texttt{gain} 
function within \xspec, which modifies the energy response according to the expression $\mathcal{E} 
= E/a + b$ (in keV). We then fitted the average 2014$m$ spectrum (obtained by merging the four 
observations) over the 1--10 keV band, by means of a phenomenological double power-law model, plus 
the required emission lines. The best slope and offset to fix the gain around the Au-M edge are 
$a \simeq 1.009$ and $b \simeq 0.004$ keV, respectively. As a by-product, these parameters also 
return sensible values of the iron-K energy scale (with the main line now falling at $E \simeq 
6.41 \pm 0.01$ keV; Section~3.2), and were thus assumed in all of the analysis. To a lesser degree, 
even the 2013 spectrum appears to be affected by some gain issues. Its response was then improved 
through a similar procedure, with $a = 1.004$ and $b = -0.002$ keV. All the line energies reported 
in the paper for both the 2013 and the 2014 \xmm observations are gain-corrected. Remarkably, as 
the 2003 EPIC/pn data are confidently exempt from calibration problems, the centroids of the two 
narrow lines in this epoch (Table~\ref{tx}) corroborate the validity of the adopted gain correction 
for the most recent spectra, independently of \chandra/HEG.

\section{Continuum modeling}
\begin{figure}
\centering
\includegraphics[width=14cm]{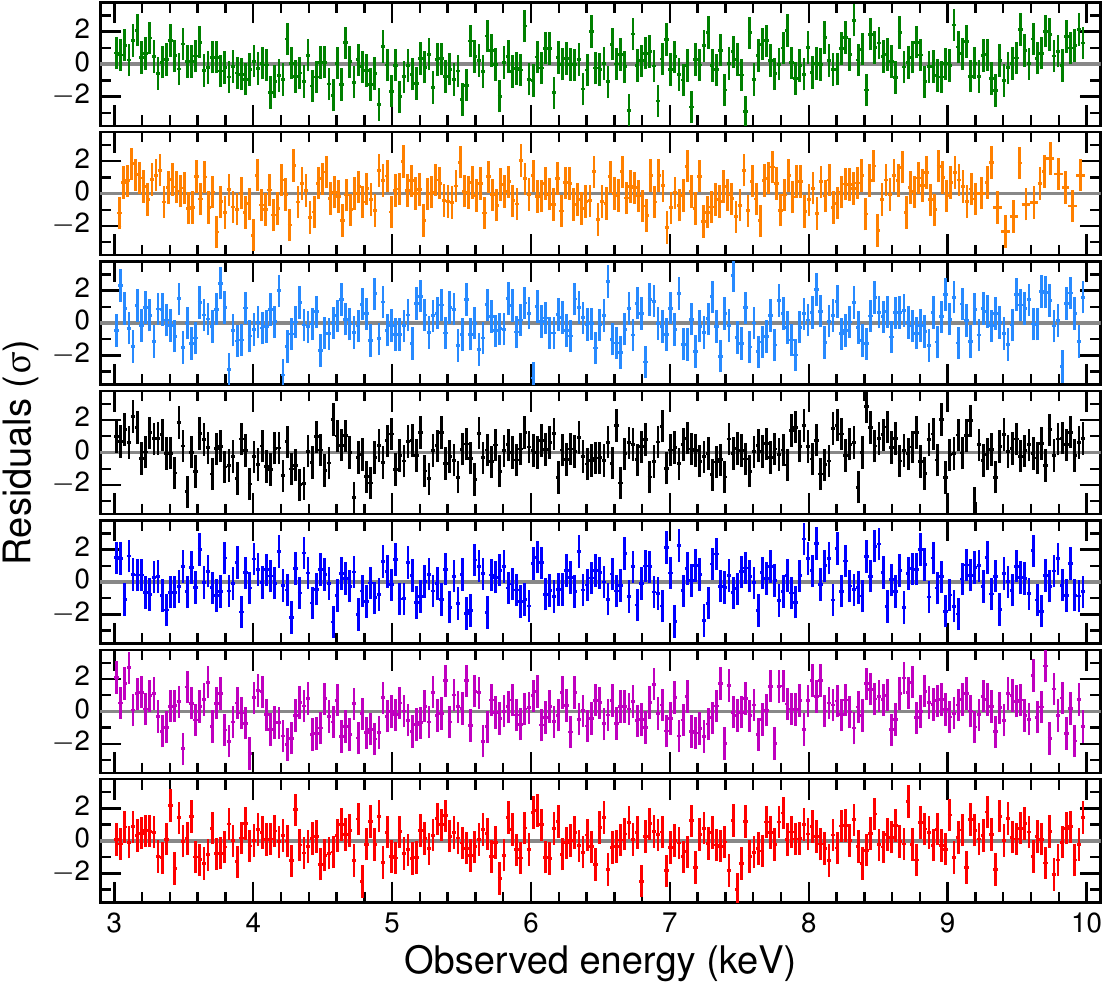}
\caption{Best-fit residuals (in units of $\sigma$; error bars have size 1) for the \relline model 
applied to all the 3--10 keV spectra of \ark (Section 4.1), showing that the continuum can be always 
well reproduced by a simple power law. In chronological order, from top to bottom: \xmm 2003, 
\suzaku 2007, \xmm 2013, 2014a, 2014b, 2014c, and 2014d (the same color code of Figures~\ref{rx} 
and \ref{ro} is used).} 
\label{re}
\end{figure}

As long as the evaluation of the emission components' parameters is the foremost intention of our 
analysis, we verified that the exact continuum shape has no practical consequence. Indeed, thanks 
to the lack of any X-ray absorption in \ark, the 3--10 keV continuum shows little complexity, and 
can be approximated with no loss of accuracy with a simple power law. For instance, here we consider 
the \relline model described in Section 4.1. By removing the three emission lines, and neglecting 
the observed 5--7 keV energy range, the fit statistic for the 3--5 plus 7--10 keV power-law continuum 
is $\chidof = 1232/1134$. Any residual spectral curvature can then be estimated by introducing a 
fictitious partial covering absorber and refitting. The improvement ($\chidof = 1191/1120$) is 
similar to the one achieved through the \relxill model (Section 3.4; $\chidof = 1196/1113$ over the 
same range), and is mainly driven by the width of the continuum window (see also footnote~3). Even 
allowing for a slight curvature, the lines' properties would be completely unaffected. The \relline 
fit residuals are shown in Figure~\ref{re}, as an illustration of the good continuum reproducibility 
through a power law. 

\section{Narrow emission from ionized iron}
Given the baffling variability properties of the apparent narrow feature at $\sim$\,7-keV, 
a logical possibility to check is whether the \kb line from neutral iron partly contributes 
to the bluest emission structures. We then introduced another Gaussian in the \relline model, 
fixing its energy to 7.06 keV and its intensity to 0.15 times the 6.4-keV \ka one (e.g., Molendi 
et al. 2003). The subsequent fit of the merged 2014$m$ \xmm spectrum proved that the new line 
is not strongly required statistically ($\dchi = -4$), and that its main effect is to push the 
neighboring, purported \fexxvi \ka redwards to 6.92$\pm$0.03 keV. Since the two features appear 
to some extent mutually exclusive, we then tried to drop the latter and let the relativistic 
line compensate for the lack of any narrow emission from ionized iron species. As a result, the 
\relline centroid moves to 6.68$^{+0.02}_{-0.03}$ keV (compatible with \fexxv \ka at rest), yet 
the red wing is not perfectly reproduced. This is reflected by the overall change in the goodness 
of fit, with $\dchidnu = 16/2$ (for an $F$-test chance probability of $1.1 \times 10^{-3}$). We 
conclude that the 7-keV excess, besides the variable component from the disk, possibly contains 
also a blend of narrow \fexxvi \ka and \fei \kb, whose relative and total intensities are very 
hard to determine. The presence of \fexxvi is supported by some conspicuous residuals at $E = 
8.20^{+0.06}_{-0.05}$ and 8.72$^{+0.06}_{-0.07}$ keV in 2014$m$, consistent with the higher-order 
\kb and K$\gamma$ transitions. When fixed at 8.25 and 8.70 keV with the same FWHM of the other 
narrow lines, these two features bring an improvement of $\dchi = 14.7$ and 15.5 against the loss 
of two d.o.f., lowering the fit statistic of 2014$m$ to $\chidof = 231/222$. Due to the huge 
difference in ionization potential, it is definitely unlikely for neutral and H-like iron to be 
co-spatial, thus requiring a substantial BLR stratification.  

\section{Variability significance}
In order to quantify the preference of a `hotspot' scenario over a standard `disk line' one, 
involving two narrow, BLR-like emission lines plus a coherent, broad feature, we also fitted the 
quiescent, red flare, and blue flare states (Figure~\ref{hs}) with the \relline model adopted 
for the time-averaged spectra. The intensity of the three lines was assumed to be constant with 
time. This returns a $\chidof$ of 529/490, clearly worse than the outcome of the hotspot model 
(500/485; with a pair of variable Gaussian/\texttt{kyr1line} components besides the 6.4-keV \ka), 
yet still broadly acceptable. With respect to the 2014$m$ best fit (Table~\ref{tx}), there are two 
main differences: the 7-keV narrow line is shifted to $E = 7.03$ keV, although the centroid energy 
is virtually unconstrained due to the sizeable intensity drop (EW $\simeq$ 10 eV); and the disc 
line is now found at $E_\rmn{R} = 6.63^{+0.17}_{-0.32}$ keV (with $r_\rmn{in} \sim 50\,\rg$). 
Overall, these findings are thus consistent with the case discussed in Appendix~C, where the 
neutral \kb takes the place of the \fexxvi \ka. As the two models are non-nested, a proper 
statistical comparision can be obtained by means of the corrected Akaike Information Criterion 
(AIC; Akaike 1974) and/or of the Bayesian Information Criterion (BIC; Schwarz 1978). For $k$ free 
parameters, $N$ data points, and $k^{\prime} = kN/(N-k-1)$, these are defined as: 
\begin{equation*}
\rmn{AIC} = \chi^2 + 2k^{\prime}, \hspace*{10pt} \rmn{BIC} = \chi^2 + k\ln N.
\end{equation*}
The correction term in AIC accounts for the finiteness of the sample. From the single AIC 
values, the Akaike weight of each model can be computed as $\mathcal{W}_\rmn{AIC} \propto 
\rmn{exp}(-\Delta_\rmn{AIC}/2)$. For the `disk line' case, it is $\mathcal{W}_\rmn{AIC} \simeq 
1.6 \times 10^{-4}$, so indicating that the `hotspot' picture is strongly preferred by the data. 
Conversely, BIC penalizes much more heavily than AIC the use of unnecessary free parameters (on 
aggregate, $16$ in the latter model against $11$ in the former), and would still marginally favor 
the `disk line' scenario in spite of the worse $\chi^2$. The difference $\Delta_\rmn{BIC} = 2$, 
however, is too small to be conclusive: the Schwarz weights, equivalent to the Akaike ones, are 
0.628 (disk line) and 0.372 (hotspot). 

\begin{figure}
\centering
\includegraphics[width=14cm]{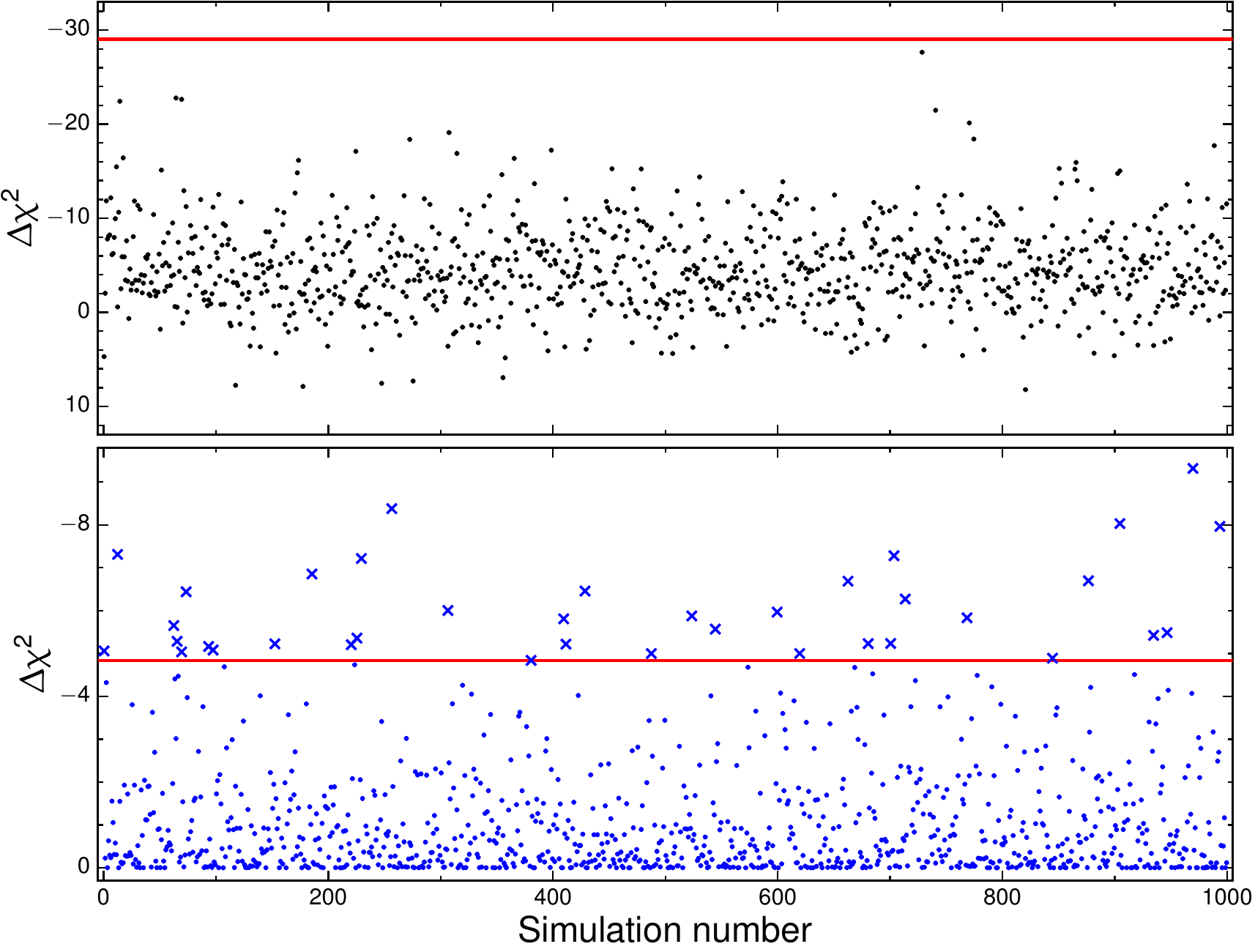}
\caption{Distributions of $\dchi$ with respect to the null-hypothesis fits obtained from Monte 
Carlo simulations. Top panel: the three fake spectra (with same exposure, background, and flux 
of the real quiescent, red flare, and blue flare states) are jointly fitted with both constant 
`disk line' and variable `hotspot' models. In a single case the statistical improvement is close 
to, yet still lower than the one found in the real data ($\dchi = -29.0$, red solid line), thus 
confirming the high significance of the short-term \fek variability. Bottom panel: the simulated 
quiescent and red flare spectra are fitted with constant (null hypothesis) and variable intensity 
of the red \fek excess. The number of false positives (37/1000) sets a conservative confidence level 
for the observed variability of the individual transients to 96.3\% (see text).} 
\label{mc}
\end{figure}

As the AIC and BIC results are contradictory, and the merits of one method over the other in any 
specific situation are still matter of debate (e.g. Liddle 2004, and references therein), we proceeded 
to further assess the true significance of \fek variability via Monte Carlo simulations. The constant 
\relline model, as evaluated above on the three distinctive spectral states extracted from the excess 
emission map, represents our null hypothesis. We first ran a test simulation, generating three spectra 
(and relative background) with the same exposure (21 ks) and response files of the quiescent, red 
flare, and blue flare states. These were jointly fitted to derive a modified null hypothesis, which 
incorporates the uncertainties due to photon noise (see also Miniutti \& Fabian 2006). With this 
refined model, we then performed the actual simulation. The output spectra were fitted twice, with 
the (constant) \relline model and the (variable) hotspot one. The $\dchi$ was recorded, and all the 
steps were repeated 1000 times. The following $\dchi$ distribution can be compared to the statistical 
improvement characterizing the observed data. In no case out of 1000 is $\dchi < -29.0$, the largest 
deviation from the null-hypothesis fit being $\dchi = -27.6$ (Figure~\ref{mc}). With a false positive 
rate of less than 10$^{-3}$, it seems highly unlikely that the red and blue \fek transient features 
are caused by noise in a constant disk-line profile. Most of all, the lack of variability is hard to 
reconcile with the quiescent phase. With this in mind, we finally checked the significance of the 
changes experienced by the red \fek excess. For simplicity, we only examined the quiescent and red 
flare states. Forcing the intensity of the red horn to be the same in both spectra, the fit deteriorates 
by $\dchi = 4.8$. We therefore ran 1000 simulations according to the procedure described above, with 
a null hypothesis of no variability. Each pair of fake data sets were then fitted with common and 
distinct normalizations of the Gaussian profile at $\sim$\,6.2 keV. The chance of obtaining a spurious 
improvement with $\dchi < -4.8$ is 37/1000 (Figure~\ref{mc}). The variability of the red feature is then 
significant at the 96.3\% level, in agreement with both the $F$-test probability (96.9\%) and the 
difference between the line's normalizations in the two spectral states ($\sim$\,2$\sigma$ confidence; 
Section~5). We note that this is a largely conservative figure from a general perspective, since the 
red excess is fainter than the blue one (maximum EW of $\sim$\,40 eV against $\sim$\,90 eV), and the 
upper limit to its intensity is even lower in the blue flare state (with $\dchi = 6.6$ for constant 
amplitude). 

%%%%%%%%%%%%%%%%%%%%%%%%%%%%%%%%%%%%%%%%%%%%%%%%%%%%%%%%%%%%%%%%%%%%%%%%%%%

%%%%%%%%%%%%%%%

\end{document}